\documentclass[11pt]{article}

\usepackage[T1]{fontenc}
\usepackage[utf8]{inputenc}

\usepackage[tt=false,type1=true]{libertine}
\usepackage[varqu]{zi4}
\usepackage[libertine]{newtxmath}
\usepackage{microtype}

\usepackage[
a4paper,
margin=1.2in,
headheight=14pt,
headsep=24pt,
footskip=30pt
]{geometry}

\usepackage{fancyhdr}
\usepackage{amsmath}
\usepackage{booktabs}
\usepackage{graphicx}
\usepackage{url}
\usepackage{hyperref}
\usepackage{xcolor}
\usepackage{lscape}
\usepackage{float}
\usepackage{makecell}
\usepackage{seqsplit}
\usepackage{array}

\newcolumntype{L}[1]{>{\raggedright\arraybackslash}p{#1}}

\usepackage[backend=biber,style=numeric]{biblatex}
\makeatletter

\newcommand{\@toptitlebar}{%
	\hrule height 2pt
	\vskip 0.25in
	\vskip -\parskip
}

\newcommand{\@bottomtitlebar}{%
	\vskip 0.29in
	\vskip -\parskip
	\hrule height 2pt
	\vskip 0.09in
}

\renewcommand{\maketitle}{%
	\par
	\begingroup
	\thispagestyle{plain}
	\begin{center}
		\vspace*{0.1in}
		\@toptitlebar
		{\LARGE\scshape \@title\par}
		\@bottomtitlebar
		\vspace{0.15in}
		{\normalsize\bfseries \@author\par}
		\vspace{0.2in}
	\end{center}
	\endgroup
}

\renewcommand{\section}{%
	\@startsection{section}{1}{\z@}%
	{-2.0ex \@plus -0.5ex \@minus -0.2ex}%
	{1.2ex \@plus 0.2ex}%
	{\large\bfseries\raggedright}}

\renewcommand{\subsection}{%
	\@startsection{subsection}{2}{\z@}%
	{-1.8ex \@plus -0.5ex \@minus -0.2ex}%
	{0.8ex \@plus 0.2ex}%
	{\normalsize\bfseries\raggedright}}

\renewcommand{\subsubsection}{%
	\@startsection{subsubsection}{3}{\z@}%
	{-1.5ex \@plus -0.5ex \@minus -0.2ex}%
	{0.5ex \@plus 0.2ex}%
	{\normalsize\bfseries\raggedright}}

\makeatother

\title{Signature Placement in Post-Quantum TLS\\Certificate Hierarchies:\\
	An Experimental Study of ML-DSA and SLH-DSA\\ in TLS 1.3 Authentication}

\author{%
	José Luis Delgado \\
	\small Universitat Oberta de Catalunya \\
	\texttt{jdelgado13@uoc.edu}
}

\begin{document}
	
	\maketitle
	
		\begin{abstract}
		Post-quantum migration in TLS 1.3 couples signature-algorithm choice with certificate-hierarchy structure, chain exposure during the handshake, and role-dependent cryptographic cost. In certificate-based authentication, the practical effect of a signature family depends on where it appears in the certification hierarchy, how much of that hierarchy is exposed during the handshake, and how the resulting cryptographic cost is distributed across client and server roles. Post-quantum TLS migration must therefore be evaluated as cryptographic design within authenticated key establishment, with algorithm selection assessed in its deployment context.
		
		This paper presents a local experimental study of TLS 1.3 authentication strategies implemented with OpenSSL~3 and oqsprovider. Using a reproducible laboratory setting, it compares ML-DSA and SLH-DSA across multiple certificate placements, hierarchy depths, and key-exchange modes, including classical, hybrid, and pure post-quantum configurations. The analysis is organized into four complementary campaigns: a leaf-only comparison, a full hierarchy strategy matrix, a depth comparison, and a key-exchange exploration.
		
		Across the experimental matrix, the main discontinuity appears when SLH-DSA is placed in the server leaf certificate. In that configuration, handshake latency and server-side compute cost increase by orders of magnitude, whereas strategies that confine SLH-DSA to upper trust layers and preserve ML-DSA in the interactive leaf remain within a more plausible operational range. The results also show that transport size alone does not explain the heavy regime: outside leaf-SLH scenarios, transferred bytes and observed chain size track latency closely, but once SLH-DSA reaches the leaf, server-side cryptographic cost becomes dominant.
		
		The paper evaluates post-quantum TLS migration as a problem of certificate-hierarchy design, chain exposure, and cryptographic cost concentration during live authentication. In practical terms, signature placement matters at least as much as signature-family choice.
	\end{abstract}
	
	
	\newpage
	\tableofcontents
	\newpage

\section{Introduction}

\subsection{Post-quantum TLS migration as protocol and PKI design}

TLS~1.3 performs authenticated key establishment through an interactive protocol exchange in which certificate transmission, signature verification, and endpoint authentication contribute to handshake cost \cite{rfc8446}. In certificate-based deployments, those costs depend on the negotiated key-establishment mechanism, the X.509 chain presented by the server, and the role of each certificate within that chain. Post-quantum migration in TLS therefore has to be studied at the level of the authenticated protocol path, where signature primitives are embedded in concrete certificate hierarchies and exercised under the timing constraints of the handshake \cite{rfc8446,sikeridis2020}.

This perspective has gained relevance after the publication of NIST's first final post-quantum standards, including ML-KEM for key establishment, ML-DSA for digital signatures, and SLH-DSA as a stateless hash-based signature standard \cite{fips203,fips204,fips205}. These standards define the primitive set considered here, but deployment choices inside latency-sensitive protocols remain protocol- and PKI-dependent. In TLS, signature families appear through certificate hierarchies, are transmitted in specific handshake messages, and are processed under interactive client/server constraints. The resulting migration problem belongs to protocol engineering and PKI design as well as to primitive selection.

The central design question in this paper concerns viability under placement. The issue is which post-quantum signature placements remain operationally plausible once TLS authentication is measured end to end, given a fixed protocol implementation and a concrete certification hierarchy. Under that reading, certificate-hierarchy design is part of the cryptographic migration problem.

\subsection{Why signature placement matters}

The cost of a signature family in TLS depends on its algorithmic properties and on its position in the certification hierarchy. A signature used in a long-lived trust anchor has a different protocol-visible effect from the same family used in the server leaf certificate presented directly in the interactive handshake. High-level discussions of post-quantum migration often obscure that distinction, even though placement is a primary deployment variable.

In TLS~1.3, the server certificate and transmitted chain participate directly in the authenticated portion of the handshake \cite{rfc8446}. The leaf certificate is therefore a special position because it binds endpoint identity to the handshake and lies at the point where certificate design, verification cost, and live server-side authentication meet. Certificates placed higher in the hierarchy may still affect validation burden, chain size, and transmitted material, but their operational role differs from that of the leaf.

A placement-sensitive study of post-quantum authentication in TLS must therefore look beyond primitive-level benchmarks and certificate-size comparisons. Those measurements remain useful, but hierarchy-sensitive placement strategies also need direct evaluation, especially when a heavier signature family is either confined to upper trust layers or placed in the handshake-exposed server leaf.

\subsection{Research gap}

Prior work has shown that post-quantum authentication in TLS can introduce substantial overhead and that certificate-related effects matter in practice \cite{sikeridis2020}. Other studies have explored mixed certificate chains as a transition strategy, making clear that heterogeneous hierarchies deserve direct analysis rather than incidental treatment as migration artifacts \cite{paul2021mixed}. More recent evaluation frameworks have expanded the measurement space by covering classical, hybrid, and pure post-quantum TLS configurations in controlled settings \cite{montenegro2025framework}.

The placement-centered question remains less studied. Existing work has often emphasized primitive-level overhead, certificate-size growth, hybrid key exchange, or aggregate TLS performance under post-quantum integration. The position of a signature family within the certificate hierarchy has received comparatively less attention, even though it shapes the effective chain observed during the handshake, the burden visible in the live authentication path, and the distribution of cryptographic work across client and server.

Hierarchy position affects which certificates are exposed, which signatures are processed in the interactive path, and where the dominant cryptographic cost is paid. A transition study that ignores placement can flatten distinct operational regimes into a single configuration-level comparison.

\subsection{Research questions and thesis}

This paper is organized around the following research questions:

\begin{enumerate}
	\item[RQ1.] To what extent is the operational cost of TLS~1.3 authentication determined by whether the server leaf certificate uses ML-DSA or SLH-DSA?
	\item[RQ2.] Does placing SLH-DSA in upper layers of the certificate hierarchy behave differently from placing it in the leaf exposed to the interactive handshake?
	\item[RQ3.] How do chain depth and effective chain transmission affect observed handshake latency and transferred data?
	\item[RQ4.] To what extent is the observed degradation explained by transport size, and to what extent by cryptographic processing?
	\item[RQ5.] Does moving from classical to hybrid or pure post-quantum key exchange materially change the main migration picture?
	\item[RQ6.] What operational implications do these results have for organizations deploying interactive TLS services?
\end{enumerate}

The paper advances a simple thesis. The dominant variable is not whether ML-DSA or SLH-DSA appears somewhere in the hierarchy, but where it appears. The main structural distinction separates scenarios in which SLH-DSA remains in upper trust layers from scenarios in which it reaches the server leaf exposed to the live TLS handshake.

The central claim follows from that distinction. Post-quantum TLS migration is best read as a certificate-hierarchy design problem. Signature placement, effective chain exposure, and client/server cryptographic burden jointly define different operational regimes, and the most severe regime is consistently associated with SLH-DSA in the interactive server leaf.

\subsection{Contributions}

\begin{enumerate}
	\item The paper presents a hierarchy-sensitive experimental study of post-quantum TLS~1.3 authentication centered on certificate placement rather than flat primitive comparison.
	\item It evaluates ML-DSA and SLH-DSA across four complementary experimental campaigns that isolate leaf placement, full hierarchy strategy, chain depth, and key-exchange mode.
	\item It provides empirical evidence that signature placement explains the observed authentication regimes better than signature-family presence alone.
	\item It shows that transport expansion does not by itself account for the dominant heavy regime observed in leaf-SLH scenarios, even though transport metrics remain highly informative outside that regime.
	\item It demonstrates, through client/server performance decomposition, that leaf-SLH scenarios become overwhelmingly server-bound during live TLS authentication.
	\item It releases the experimental code and scenario-generation material in a public repository to support traceability and reproducibility: \href{https://github.com/hypergalois/PKI-PQC}{https://github.com/hypergalois/PKI-PQC}.
	\item It translates the measured effects into operational terms, connecting post-quantum certificate-hierarchy design with deployment viability in interactive TLS services.
\end{enumerate}

\subsection{Paper organization}

Section~2 introduces the cryptographic and protocol background needed to frame post-quantum authentication in TLS~1.3. Section~3 reviews the most relevant prior work and positions the present study within that literature. Section~4 states the research questions more formally and defines the scope of the study. Section~5 presents the experimental methodology, including the implementation stack, scenario construction, measurement model, sampling strategy, and metric semantics. Section~6 reports the campaign-level results. Section~7 develops a cross-cutting cryptographic interpretation of those results, with particular attention to signature placement, chain exposure, and the distinction between transport-related and cryptographic cost. Section~8 analyzes client/server workload decomposition. Section~9 translates the measured effects into operational implications for post-quantum TLS deployment. Section~10 discusses threats to validity and limitations. Section~11 concludes.

\section{Cryptographic and Protocol Background}

\subsection{TLS 1.3 certificate authentication and chain transmission}

TLS~1.3 provides authenticated key establishment by combining ephemeral key exchange with transcript-bound authentication \cite{rfc8446}. In the certificate-based server path, the server sends a certificate chain in the \texttt{Certificate} message and proves possession of the corresponding private key in \texttt{CertificateVerify}. Both messages are bound to the handshake transcript, so certificate handling belongs to the live cryptographic execution of the protocol and its cost is observed during the handshake.

In Internet deployments, the certificates used along that path typically follow the X.509 profile standardized for PKIX, with a leaf certificate identifying the authenticated endpoint and additional certificates carrying the issuing chain required for validation \cite{rfc5280}. TLS authentication cost therefore depends on the negotiated key-establishment mechanism, the structure and contents of the certificate path presented during the handshake, and the position of each certificate in that path. The server leaf has a special role because it is tied directly to endpoint identity and to the key used in the interactive authentication step.

TLS authenticates a concrete endpoint through a concrete certificate path transmitted under protocol timing constraints. Certificate hierarchy, effective chain exposure, and verification burden can therefore affect the observed behavior of the handshake. In the post-quantum setting, where signature families may differ sharply in size and computational profile, those effects become more visible \cite{sikeridis2020,paul2021mixed}.

\subsection{Post-quantum signature standardization: ML-DSA and SLH-DSA}

NIST's first final post-quantum standards define ML-KEM for key establishment, ML-DSA for digital signatures, and SLH-DSA as a stateless hash-based digital signature standard \cite{fips203,fips204,fips205}. For TLS authentication, ML-DSA and SLH-DSA are standardized signature options with different design foundations within post-quantum cryptography.

ML-DSA is a module-lattice-based signature standard derived from the CRYSTALS-Dilithium lineage and intended as a general-purpose digital signature mechanism for the post-quantum transition \cite{fips204}. SLH-DSA is a stateless hash-based standard derived from the SPHINCS+ family and reflects a different set of cryptographic trade-offs grounded in hash-based security assumptions and stateless tree-based signing \cite{fips205}. Both are standardized options, but standardization does not imply similar behavior once they are embedded in certificate hierarchies and exercised inside a live TLS handshake.

The present paper keeps that distinction at the level relevant to the protocol setting under study. It does not attempt a full algorithmic comparison of ML-DSA and SLH-DSA across all possible uses. Its narrower concern is TLS~1.3 certificate authentication, where signature families are consumed through transmitted and validated certification paths. In that setting, standardization establishes cryptographic legitimacy, but not operational equivalence.

\subsection{Hybrid transition logic and deployment tension}

The transition to post-quantum cryptography is widely understood as a migration problem rather than a one-step replacement event. In TLS, this has been especially visible on the key-establishment side, where hybrid designs combine classical and post-quantum components so that security is preserved if at least one constituent remains secure during the transition period \cite{turner2026hybrid,mlkemtlsdraft}. That logic is motivated by compatibility, cryptographic conservatism, and the practical need to move incrementally instead of assuming immediate ecosystem-wide convergence \cite{uta2026pqcapp}.

A related tension appears in certificate-based authentication. Organizations migrate certificate hierarchies under trust-anchor lifetimes, interoperability constraints, implementation limits, validation behavior, and operational latency budgets. In TLS, the post-quantum transition is therefore tied to the choice of signature primitives, to how those primitives are distributed across the certification hierarchy, and to which costs are exposed in the live handshake \cite{paul2021mixed,sikeridis2020}.

Mixed and hierarchy-sensitive strategies address that deployment setting directly. A deployment may place a heavier or more conservative signature family in upper trust layers while preserving a different family at the interactive server leaf, or it may choose the reverse placement. Those choices are part of transition design because they determine how authentication cost appears during protocol execution.

\subsection{Placement, exposure, and operational cost}

The central premise of this paper is that certificate placement within the hierarchy is a primary variable in post-quantum TLS authentication. The relevant issue is which signature family appears in the chain, which certificates are exposed during the handshake, and where the corresponding cryptographic burden is paid. Prior work on post-quantum authentication in TLS and on mixed certificate chains suggests that certificate-related effects can dominate practical overhead and that heterogeneous hierarchies require explicit treatment during the transition \cite{sikeridis2020,paul2021mixed}.

That point becomes sharper once logical hierarchy is separated from effective chain exposure. A PKI may contain roots, intermediates, and leaves as part of its declared structure, but the set of certificates actually transmitted and processed during the handshake is an empirical property of the deployed authentication path. Effective exposure influences transmitted bytes, validation burden, and potentially the concentration of active work across client and server roles. It is therefore part of the cryptographic cost surface of the protocol rather than a secondary representational detail.

Under that view, placement has direct operational meaning. A signature family used in an upper trust layer may affect path validation and chain size while leaving the live server-authentication step within a bounded cost regime. The same family, once moved into the handshake-exposed leaf, may alter the execution profile of the entire authentication path. The present study examines that distinction systematically by treating hierarchy position, effective chain exposure, and client/server cost concentration as joint variables in post-quantum TLS~1.3 authentication.

\section{Related Work}

\subsection{Prior work on post-quantum TLS and hybrid handshakes}

Prior work has established TLS as a main protocol setting for empirical evaluation of post-quantum migration beyond primitive-level analysis. Sikeridis, Kampanakis, and Devetsikiotis studied post-quantum authentication in TLS~1.3 and showed that signature choice can materially affect connection-establishment latency and throughput under realistic deployment assumptions \cite{sikeridis2020}. Their results quantify overhead in the authentication path and show that this path is a relevant deployment surface for post-quantum cryptography.

Subsequent work expanded the transition picture by examining mixed and hybrid configurations. In the authentication space, Paul et al.\ proposed mixed certificate chains for TLS~1.3 as an explicit migration strategy, arguing that different signature algorithms may be assigned to different positions in the same hierarchy to balance trust, interoperability, and performance \cite{paul2021mixed}. On the key-establishment side, current IETF work on hybrid key exchange and ML-KEM deployment in TLS follows a related transition model, in which compatibility and security margin have to coexist during staged deployment \cite{turner2026hybrid,mlkemtlsdraft}.

The practical experimentation ecosystem has also matured. The Open Quantum Safe project is a common platform for prototyping post-quantum cryptography in real protocol stacks, including OpenSSL-based TLS deployments through liboqs and oqsprovider \cite{oqs,oqsprovider}. More recent measurement-oriented work, such as the framework of Montenegro et al., has expanded the empirical basis for post-quantum TLS evaluation by enabling comparative study across classical, hybrid, and pure post-quantum configurations in a unified environment \cite{montenegro2025framework}. Taken together, these lines of work define PQ TLS as a concrete domain of empirical cryptographic research.

\subsection{Prior work on certificate overhead and chain-related cost}

Within that literature, certificate-related overhead has emerged as a major source of deployment cost. Sikeridis et al.\ showed that post-quantum authentication in TLS~1.3 cannot be evaluated solely through abstract security arguments or primitive-level operation counts, because certificate size, transmitted bytes, and signature verification cost all affect handshake behavior materially \cite{sikeridis2020}. Their study shifted the discussion from purely algorithmic comparison toward protocol-visible authentication overhead.

Paul et al.\ extended that line of analysis by treating the certificate chain itself as a design space instead of a fixed administrative artifact \cite{paul2021mixed}. Their mixed-chain proposal is especially relevant because it recognizes that post-quantum migration may proceed through heterogeneous hierarchies in which different certificates serve different operational roles. It suggests that the effective cost of post-quantum authentication depends on algorithm selection and on the distribution of those algorithms across the hierarchy.

Recent PQ TLS evaluation frameworks have continued to track the relation between cryptographic choices, handshake latency, and protocol overhead \cite{montenegro2025framework}. Even so, prior work has generally emphasized aggregate configuration-level behavior more than placement-centered analysis of certification hierarchies. The literature has shown that certificates matter. It has been less explicit about how hierarchy position, effective chain exposure, and client/server cost concentration interact as distinct explanatory variables.

\subsection{Positioning of this paper}

The present study addresses that placement-centered gap. It focuses on the certificate hierarchy itself: which signature family is placed at the root, which at the intermediate, which at the server leaf, and how that placement shapes the cost of live TLS authentication. The study includes classical, hybrid, and pure post-quantum KEX modes in the experimental design, but its main object is the authentication hierarchy rather than key-exchange substitution or uniform signature replacement.

This choice distinguishes the present work from prior evaluation efforts in two ways. First, hierarchy-sensitive signature placement is treated as the main explanatory variable rather than as a secondary parameter inside a broader benchmark. Second, hierarchy placement is analyzed together with effective chain exposure and client/server workload decomposition during the handshake. The paper studies post-quantum TLS authentication as a problem of certificate-path construction and exposure, as well as aggregate overhead \cite{sikeridis2020,paul2021mixed,montenegro2025framework}.

The paper asks which certificate-chain strategies remain operationally plausible once post-quantum signature placement is evaluated inside the authenticated handshake path of TLS~1.3.

\section{Research Questions and Study Scope}

\subsection{Research questions}

The study is organized around six research questions:

\begin{enumerate}
	\item[RQ1.] To what extent is the operational cost of TLS~1.3 certificate-based authentication determined by whether the server leaf certificate uses ML-DSA or SLH-DSA?
	
	\item[RQ2.] Does placing SLH-DSA in upper layers of the certificate hierarchy behave materially differently from placing it in the handshake-exposed server leaf?
	
	\item[RQ3.] How do hierarchy depth and effective chain exposure shape observed handshake latency and transmitted data during TLS~1.3 authentication?
	
	\item[RQ4.] To what extent is the observed degradation explained by transport-related overhead, and to what extent by cryptographic processing cost?
	
	\item[RQ5.] Does moving from classical to hybrid or pure post-quantum key establishment materially alter the main migration picture?
	
	\item[RQ6.] What operational implications do these results have for organizations deploying interactive TLS services under post-quantum transition constraints?
\end{enumerate}

Taken together, these questions define the analytical focus of the paper. The study determines how certificate-hierarchy design shapes the cost of live TLS~1.3 authentication once signature placement, chain exposure, and client/server burden are considered jointly.

\subsection{Scope and non-goals}

The paper studies a concrete cryptographic and operational question: how post-quantum signature placement within TLS~1.3 certificate hierarchies affects the behavior of live certificate-based authentication in an implementation stack built on OpenSSL~3 and oqsprovider \cite{rfc8446,oqs,oqsprovider}. The aim is controlled evaluation of a specific migration problem with practical relevance, not universal prediction across all TLS libraries, PKI deployments, or Internet environments.

The scope is bounded in several ways. First, the evaluation targets one implementation stack. Absolute values may depend on library internals, provider integration, certificate handling behavior, and platform-specific factors. The strongest claims advanced here are structural: hierarchy-sensitive signature placement can decisively shape the operational regime of post-quantum TLS authentication.

Second, the analysis does not include full internal function tracing or exhaustive microarchitectural attribution. Performance counters are used as regime-level evidence for client/server cost concentration, not as a replacement for complete code-path reconstruction or low-level implementation forensics.

Third, the study does not measure Internet-wide or WAN-visible latency under arbitrary production conditions. The experiments are conducted in a controlled local environment so that certificate-path effects, chain exposure, and cryptographic cost can be observed with reduced network noise. The reported latency values should therefore be read as comparative measurements within a real stack, not as direct forecasts of user-perceived latency on the public Internet.

Finally, the study does not settle which post-quantum primitive is best across all protocol settings. It addresses a narrower question: which certificate-chain strategies remain operationally plausible for TLS~1.3 server authentication when post-quantum signatures are embedded into live X.509-based authentication paths \cite{rfc8446,rfc5280,fips204,fips205}.

\section{Experimental Methodology}

\subsection{Experimental objective}

The experimental objective is to measure how the placement of ML-DSA and SLH-DSA within X.509 certification paths affects the cost of live server authentication during the TLS handshake \cite{rfc8446,rfc5280,fips204,fips205}.

The laboratory evaluates hierarchy-sensitive authentication scenarios at the protocol level. The study asks which hierarchy constructions remain operationally plausible once authentication is exercised through a concrete certificate path, transmitted during a real TLS~1.3 handshake, and measured as an end-to-end cryptographic event. This framing is consistent with prior work showing that post-quantum authentication overhead in TLS is shaped by certificate-related effects and is best studied at the protocol level instead of being inferred from primitive properties alone \cite{sikeridis2020,paul2021mixed,montenegro2025framework}.

\subsection{Implementation stack and execution environment}

All experiments were conducted in a local TLS laboratory built around OpenSSL~3 and the Open Quantum Safe provider stack. The cryptographic implementation used oqsprovider on top of liboqs, which enabled the post-quantum signature and key-establishment mechanisms required for the study \cite{oqs,oqsprovider}. The signature families of interest were ML-DSA and SLH-DSA, instantiated through concrete algorithm selections appropriate to the chosen security level. On the key-establishment side, the experimental design considered three modes: classical \texttt{X25519}, hybrid \texttt{X25519MLKEM768}, and pure post-quantum \texttt{MLKEM768} \cite{fips203,fips204,fips205,mlkemtlsdraft,turner2026hybrid}.

The server side was driven through \texttt{openssl s\_server}, parameterized per scenario to load the corresponding certificate chain, key material, and TLS group configuration. The client side used a dedicated benchmark client written in C and instrumented for per-handshake measurement. For each run, the client established a fresh TCP connection, forced TLS~1.3, loaded the required provider configuration, selected the key-establishment group for the scenario under test, and executed a full handshake against the server. The unit of observation is therefore the complete certificate-based authentication event, excluding resumed sessions and amortized multi-request connection artifacts.

The code used to generate certificates, construct scenarios, drive measurements, and organize the experimental artifacts is available in the public repository \href{https://github.com/hypergalois/PKI-PQC}{https://github.com/hypergalois/PKI-PQC}.

\subsection{Scenario construction}

The scenario space was designed to isolate hierarchy-sensitive authentication effects. Each scenario is defined by a combination of certificate-hierarchy depth, signature-family placement within the hierarchy, and key-establishment mode.

Hierarchy depth was varied between depth~2 and depth~3 configurations. In depth~2 scenarios, the logical certification path consisted of a root and a leaf. In depth~3 scenarios, an intermediate certificate was inserted between root and leaf. This distinction was introduced to examine how hierarchy depth interacts with effective chain exposure and with the placement of heavier signature families.

Within each hierarchy, signature families were assigned positionally. The main placement variables are therefore root, intermediate, and leaf. This positional model distinguishes scenarios in which SLH-DSA appears only in upper trust layers, scenarios in which it appears directly in the server leaf, and scenarios in which the entire hierarchy uses a common family. The design supports the question of whether the same signature family has different operational meaning depending on where it is embedded in the certification path.

The experimental matrix included both uniform and mixed hierarchies. Uniform hierarchies were used to define the all-ML baseline and selected all-SLH comparison points. Mixed hierarchies were used to evaluate transitional strategies such as root-SLH with ML-DSA preserved in the leaf, as well as heavier placements in which SLH-DSA reaches the interactive certificate. The key-establishment dimension was varied across classical, hybrid, and pure post-quantum modes to determine whether the placement-sensitive picture changes under different KEX assumptions.

For analysis purposes, the study also defines comparable families of scenarios. These groups differ along one principal axis while preserving the rest of the hierarchy structure as far as possible. Examples include leaf-only contrasts under fixed surrounding conditions, depth~2 versus depth~3 comparisons within the same high-level family, and KEX comparisons under comparable chain constructions. Comparability supports causal interpretation because the paper is concerned less with aggregate scenario ranking than with identifying which structural variable governs each observed regime.

\subsection{Measurement model}

The measurement model is defined at the level of individual handshakes. Each experimental run establishes a fresh TCP connection, performs a TLS~1.3 handshake under the selected scenario, and records protocol-visible and system-visible metrics. The per-handshake design uses live certificate-based authentication as the basic unit of analysis.

Handshake latency was measured using a monotonic clock source to avoid wall-clock artifacts and preserve a stable basis for repeated comparisons. At the transport level, the client recorded the number of bytes read and written during the handshake. At the certificate-path level, it recorded the observed chain length and the observed chain size in DER bytes, using the peer certificate chain exposed by the TLS stack together with certificate serialization. A defensive fallback to the leaf certificate was maintained so that the measurement pipeline continued to produce data when chain exposure did not match a naive logical expectation.

In addition to protocol-level metrics, the experiments collected performance-counter data on both client and server. These counters were normalized per run to support client/server decomposition at the handshake level. The principal performance-counter variables used in the paper are task-clock, instructions, cycles, and derived ratios such as task-clock over elapsed time and server/client task-clock ratio. These variables indicate whether a given regime remains balanced, becomes validation-skewed, or becomes overwhelmingly server-bound.

\subsection{Sample size, measurement stability, and interpretive scope}

Run counts were not uniform across scenarios. Fast scenarios were executed with larger sample sizes, whereas extremely heavy scenarios, especially those involving SLH-DSA in the leaf, were run with smaller but still analytically sufficient counts. When one class of scenarios completes in well under a millisecond and another takes on the order of seconds, enforcing identical run counts across the entire matrix would spend disproportionate wall-clock budget on scenarios whose regime is already sharply separated from the rest.

This choice should be read in light of the inferential aim of the paper. The study is not trying to resolve tiny marginal differences among near-equivalent configurations. Its main objective is to identify regime changes, placement-sensitive discontinuities, and shifts in where cryptographic work is concentrated. Under that objective, a smaller number of runs in a seconds-scale scenario does not carry the same interpretive risk that it would in a study focused on sub-percent differences between tightly clustered alternatives.

The non-uniform sampling policy balances feasibility and descriptive stability. Fast scenarios were run more extensively because they were inexpensive to execute and because denser repetition is useful when characterizing low-latency regions. Heavy leaf-SLH scenarios were run fewer times because each run was costly, yet the observed differences remained large, consistent, and visible across multiple metrics, including mean latency, p95 latency, transport volume, and server-side task-clock.

\subsection{Experimental campaigns}

The experimental design is organized into four campaigns, each isolating a distinct question in post-quantum TLS authentication.

Campaign~A isolates the leaf effect. It compares scenarios in which the principal change is the signature family used in the server leaf, while keeping the surrounding hierarchy as simple and controlled as possible. Its purpose is to test whether the main discontinuity already appears in the most direct certificate-authentication comparison.

Campaign~B studies the full hierarchy strategy space under a common hybrid key-establishment regime. This is the central strategy matrix of the paper. It evaluates complete hierarchy constructions rather than local pairwise substitutions, allowing the study to compare migration strategies as complete authentication designs.

Campaign~C isolates topology from placement by comparing depth~2 and depth~3 hierarchies across comparable algorithmic families. Its purpose is to determine whether increasing chain depth behaves as a simple monotonic penalty or whether the effect depends on which certificates become visible in the effective chain transmitted during the handshake.

Campaign~D explores key-establishment variation under comparable chain constructions. It contrasts classical, hybrid, and pure post-quantum KEX modes to determine whether the main placement-driven interpretation survives across different key-establishment assumptions.

\subsection{Metric semantics and normalization}

The paper uses a layered metric model. At the handshake level, the main latency variables are \texttt{elapsed\_mean\_ms} and \texttt{elapsed\_p95\_ms}, which capture average and high-percentile latency, respectively. These are the principal observables used to assess interactive plausibility in live TLS authentication.

At the transport and chain level, the main variables are \texttt{bytes\_read\_mean}, \texttt{bytes\_written\_mean}, \texttt{chain\_bytes\_unique}, and \texttt{served\_chain\_der\_bytes}. The first two track protocol-visible byte cost during the handshake. The latter two characterize certificate material exposure more directly, although their semantics are not perfectly homogeneous across all topology classes, as discussed below.

At the performance-counter level, the main variables are \texttt{client\_task\_clock\_per\_run\_ms}, \texttt{server\_task\_clock\_per\_run\_ms}, instructions, cycles, IPC, and derived normalized ratios. These variables allow the analysis to distinguish among balanced low-cost regimes, client-skewed validation regimes, and overwhelmingly server-bound regimes.

Several metrics are also expressed relative to a common baseline. The baseline scenario used throughout the paper is the hybrid depth~3 fully-ML hierarchy:
\[
\texttt{x25519mlkem768\_\_ml\_root\_\_ml\_int\_\_ml\_leaf}.
\]
Relative metrics such as latency multiplier, bytes-read multiplier, retained capacity, and server-side compute multiplier make structural contrasts easier to read than raw absolute values alone. This normalization is especially useful because the experimental space contains both sub-millisecond and second-scale regimes.

Finally, the paper derives capacity and economic interpretations from normalized server-side compute measurements. These derived quantities are not treated as direct billing observations, but as operational translations of per-handshake server burden.

\subsection{Methodological cautions}

Several methodological cautions are necessary for correct interpretation of the dataset.

First, the analytical dataset used for cross-cutting interpretation is deduplicated by \texttt{scenario\_id}. Campaign~C intentionally reuses scenarios that also appear in Campaign~B. These repeated scenarios were checked for metric consistency and treated as analytical reuse rather than as independent new observations. Campaign identity is preserved where narratively relevant, but scenario-level cross-cutting analysis operates on the deduplicated space.

Second, the observed value \texttt{chain\_len\_unique = 2} does not imply that the logical certification hierarchy has depth~2. In the dataset studied here, the client consistently observes two certificates in the effective chain, but those two certificates do not always correspond to the same logical pair. In depth~2 scenarios, the observed pair may correspond to root plus leaf; in depth~3 scenarios, it may instead correspond to intermediate plus leaf. Effective chain exposure must therefore be treated as an empirical variable rather than inferred mechanically from declared hierarchy depth.

Third, \texttt{served\_chain\_der\_bytes} does not carry identical semantics across all topology classes. In the present dataset, it aligns with leaf DER size in some depth~2 scenarios, whereas in depth~3 scenarios it more closely tracks the observed effective chain. For strict transport reasoning across depths, the more reliable observables are therefore \texttt{bytes\_read\_mean} and \texttt{chain\_bytes\_unique}.

Fourth, the performance-counter analysis is regime-oriented. The paper uses perf-derived data to identify where active work concentrates during the handshake, not to provide complete code-path attribution. Strong claims are therefore made at the level of cost concentration and regime structure, rather than at the level of unique internal implementation causes.

\section{Results}

\subsection{Global performance landscape}

The global results separate into distinct operational regimes rather than a smooth continuum of gradually increasing cost. All-ML scenarios remain in a narrow low-latency band below one millisecond, whereas scenarios with SLH-DSA in the server leaf cluster around a plateau near 1.4 seconds. Between these regions lies a smaller intermediate band, populated mainly by configurations that place SLH-DSA in upper hierarchy layers while preserving ML-DSA in the interactive leaf.

\begin{figure}[t]
	\centering
	\includegraphics[width=\linewidth]{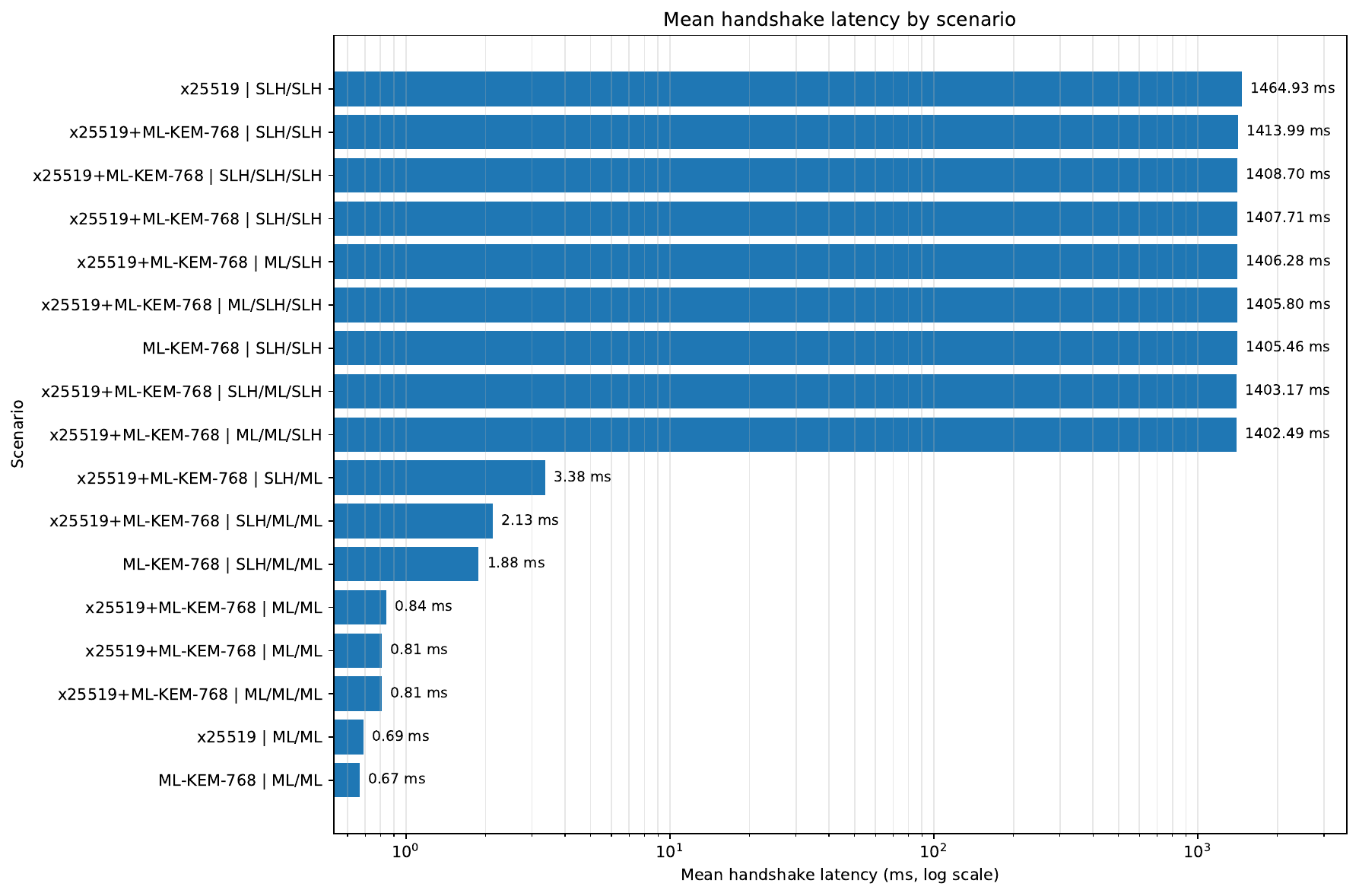}
	\caption{Mean handshake latency by scenario.}
	\label{fig:fig01_mean_latency_by_scenario}
\end{figure}

\begin{figure}[t]
	\centering
	\includegraphics[width=\linewidth]{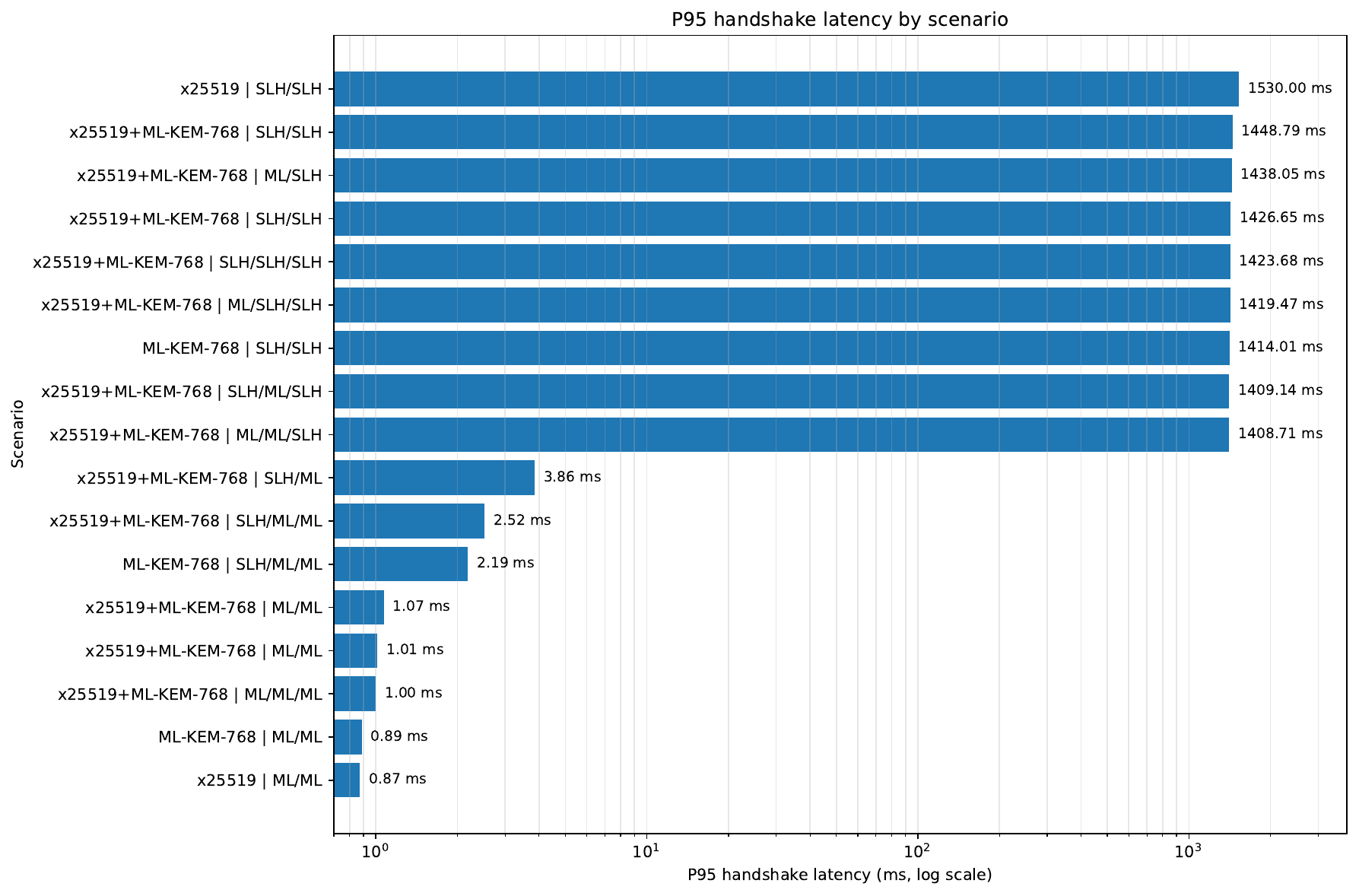}
	\caption{P95 handshake latency by scenario.}
	\label{fig:fig02_p95_latency_by_scenario}
\end{figure}

\begin{table}[t]
	\centering
	\caption{Master scenario summary (short version).}
	\label{tab:block6_table_master_summary_short}
	\resizebox{\textwidth}{!}{%
		\begin{tabular}{lllllllll}
\toprule
scenario\_id & kex\_mode & depth & hierarchy & mean\_ms & p95\_ms & bytes\_read & server\_task\_ms & srv\_cli\_ratio \\
\midrule
x25519\_\_leaf\_mldsa65 & classical & 2 & ML root / ML leaf & 0.688 & 0.874 & 14904 & 0.529 & 1.022 \\
x25519\_\_leaf\_slhdsashake192s & classical & 2 & SLH root / SLH leaf & 1464.933 & 1529.999 & 49881 & 1462.281 & 462.869 \\
x25519mlkem768\_\_leaf\_mldsa65 & hybrid & 2 & ML root / ML leaf & 0.841 & 1.072 & 15992 & 0.623 & 1.049 \\
x25519mlkem768\_\_leaf\_slhdsashake192s & hybrid & 2 & SLH root / SLH leaf & 1413.991 & 1448.787 & 50969 & 1411.345 & 448.264 \\
x25519mlkem768\_\_ml\_root\_\_ml\_leaf & hybrid & 2 & ML root / ML leaf & 0.809 & 1.013 & 15992 & 0.602 & 1.044 \\
x25519mlkem768\_\_ml\_root\_\_slh\_leaf & hybrid & 2 & ML root / SLH leaf & 1406.283 & 1438.054 & 26999 & 1404.927 & 696.186 \\
x25519mlkem768\_\_slh\_root\_\_ml\_leaf & hybrid & 2 & SLH root / ML leaf & 3.376 & 3.856 & 39962 & 1.915 & 0.998 \\
x25519mlkem768\_\_slh\_root\_\_slh\_leaf & hybrid & 2 & SLH root / SLH leaf & 1407.714 & 1426.652 & 50969 & 1405.087 & 441.560 \\
x25519mlkem768\_\_ml\_root\_\_ml\_int\_\_ml\_leaf & hybrid & 3 & ML root / ML int / ML leaf & 0.809 & 1.000 & 16008 & 0.562 & 0.903 \\
x25519mlkem768\_\_ml\_root\_\_ml\_int\_\_slh\_leaf & hybrid & 3 & ML root / ML int / SLH leaf & 1402.486 & 1408.711 & 27015 & 1401.169 & 687.949 \\
x25519mlkem768\_\_ml\_root\_\_slh\_int\_\_slh\_leaf & hybrid & 3 & ML root / SLH int / SLH leaf & 1405.803 & 1419.470 & 38046 & 1403.109 & 433.616 \\
x25519mlkem768\_\_slh\_root\_\_ml\_int\_\_ml\_leaf & hybrid & 3 & SLH root / ML int / ML leaf & 2.133 & 2.522 & 28947 & 0.667 & 0.346 \\
x25519mlkem768\_\_slh\_root\_\_ml\_int\_\_slh\_leaf & hybrid & 3 & SLH root / ML int / SLH leaf & 1403.166 & 1409.142 & 39954 & 1401.849 & 430.764 \\
x25519mlkem768\_\_slh\_root\_\_slh\_int\_\_slh\_leaf & hybrid & 3 & SLH root / SLH int / SLH leaf & 1408.703 & 1423.678 & 50985 & 1404.903 & 326.456 \\
mlkem768\_\_ml\_root\_\_ml\_leaf & pure\_pqc & 2 & ML root / ML leaf & 0.665 & 0.887 & 15960 & 0.515 & 1.003 \\
mlkem768\_\_slh\_root\_\_slh\_leaf & pure\_pqc & 2 & SLH root / SLH leaf & 1405.456 & 1414.012 & 50937 & 1402.896 & 463.910 \\
mlkem768\_\_slh\_root\_\_ml\_int\_\_ml\_leaf & pure\_pqc & 3 & SLH root / ML int / ML leaf & 1.884 & 2.186 & 28915 & 0.527 & 0.298 \\
\bottomrule
\end{tabular}

	}
\end{table}

Figure~\ref{fig:fig01_mean_latency_by_scenario} shows mean latency ranging from approximately 0.665~ms to 1464.933~ms. This spread reflects regime separation rather than a modest gradient of penalties. The low-cost region contains all-ML scenarios and a small number of root-SLH / leaf-ML cases, while the heavy region is dominated by configurations in which SLH-DSA reaches the leaf.

Figure~\ref{fig:fig02_p95_latency_by_scenario} shows the same structure in high-percentile behavior. The heavy scenarios shift upward in both mean and upper-tail latency, indicating that the severe regime is a persistent property of the handshake rather than a minor average-case distortion.

The heavy scenarios do not follow a broad gradient in which progressively larger chains produce progressively worse latency. Once SLH-DSA is placed in the leaf, they cluster into a relatively tight plateau, which points to placement as the main discontinuity rather than to incremental transport growth alone.

\subsection{Campaign A: leaf-only comparison}

Campaign~A isolates the leaf effect while keeping the surrounding hierarchy simple and controlled. It tests whether the main discontinuity appears as soon as SLH-DSA is introduced into the interactive server leaf.

\begin{figure}[t]
	\centering
	\includegraphics[width=0.9\linewidth]{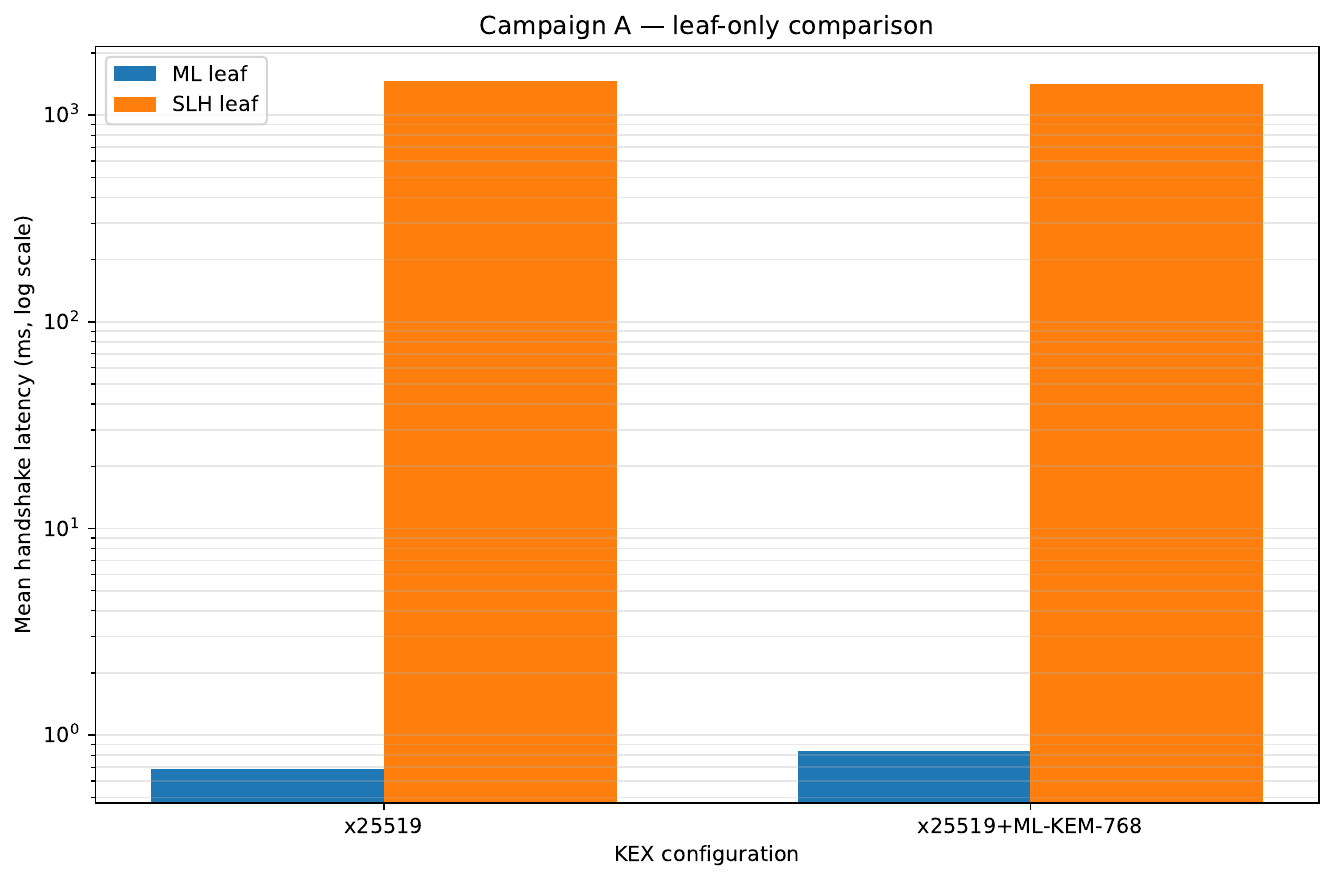}
	\caption{Campaign A: leaf-only comparison.}
	\label{fig:fig03_campaignA_leaf_only_comparison}
\end{figure}

\begin{table}[t]
	\centering
	\caption{Campaign A paired comparison.}
	\label{tab:block7A_campaignA_pairwise_ratios_body}
	\resizebox{\textwidth}{!}{%
	\begin{tabular}{llrrrrrr}
		\toprule
		kex\_mode &
		tls\_group\_family &
		\makecell[r]{ml elapsed\\mean (ms)} &
		\makecell[r]{slh elapsed\\mean (ms)} &
		\makecell[r]{latency ratio\\SLH / ML} &
		\makecell[r]{ml bytes\\read mean} &
		\makecell[r]{slh bytes\\read mean} &
		\makecell[r]{bytes-read ratio\\SLH / ML} \\
		\midrule
		classical & x25519 & 0.688 & 1464.933 & 2127.865 & 14904 & 49881 & 3.347 \\
		hybrid & x25519mlkem768 & 0.841 & 1413.991 & 1682.137 & 15992 & 50969 & 3.187 \\
		\bottomrule
	\end{tabular}%
}

\vspace{0.5em}

\resizebox{\textwidth}{!}{%
	\begin{tabular}{llrrrrrr}
		\toprule
		kex\_mode &
		tls\_group\_family &
		\makecell[r]{ml server task\\clock per run (ms)} &
		\makecell[r]{slh server task\\clock per run (ms)} &
		\makecell[r]{server taskclock\\ratio SLH / ML} &
		\makecell[r]{ml client task\\clock per run (ms)} &
		\makecell[r]{slh client task\\clock per run (ms)} &
		\makecell[r]{client taskclock\\ratio SLH / ML} \\
		\midrule
		classical & x25519 & 0.529 & 1462.281 & 2765.628 & 0.517 & 3.159 & 6.107 \\
		hybrid & x25519mlkem768 & 0.623 & 1411.345 & 2265.961 & 0.594 & 3.148 & 5.303 \\
		\bottomrule
	\end{tabular}%
}
\end{table}

\subsubsection{Classical KEX}

Under classical \texttt{X25519}, replacing an ML-DSA leaf with an SLH-DSA leaf produces an extreme discontinuity. Mean latency increases by approximately 2127.86$\times$, and p95 by approximately 1750.57$\times$. The transport expansion is much smaller: \texttt{bytes\_read\_mean} grows by about 3.35$\times$, and \texttt{chain\_bytes\_unique} by about 2.98$\times$.

The same asymmetry appears in the performance counters. Server task-clock per run grows by approximately 2765.63$\times$, whereas client task-clock per run grows by only 6.11$\times$. Even in this simplest leaf-only configuration, the latency jump is far larger than the increase in transmitted certificate material.

\subsubsection{Hybrid KEX}

The same comparison under hybrid \texttt{X25519MLKEM768} yields the same qualitative result. Mean latency increases by approximately 1682.14$\times$, while p95 rises by approximately 1351.48$\times$. Transport growth again remains comparatively modest: \texttt{bytes\_read\_mean} grows by about 3.19$\times$, and \texttt{chain\_bytes\_unique} by about 2.98$\times$.

The performance counters support the same interpretation. Server task-clock per run increases by approximately 2265.96$\times$, while client task-clock per run increases by only 5.30$\times$. The heavy regime is therefore visible already in the simplest leaf-only comparison and does not depend on deeper hierarchies or more elaborate mixed chains.

\subsubsection{Main result of Campaign~A}

Campaign~A shows that the dominant operational discontinuity does not require a complex hierarchy, a mixed certification strategy, or a pure post-quantum key-establishment path. It appears as soon as SLH-DSA is moved into the server leaf certificate, before the analysis reaches the full hierarchy design space.

\subsection{Campaign B: full hierarchy strategy matrix}

Campaign~B evaluates complete hierarchy constructions under a common hybrid key-establishment regime. This campaign compares migration strategies as full authentication designs rather than as local substitutions.

\begin{figure}[t]
	\centering
	\includegraphics[width=\linewidth]{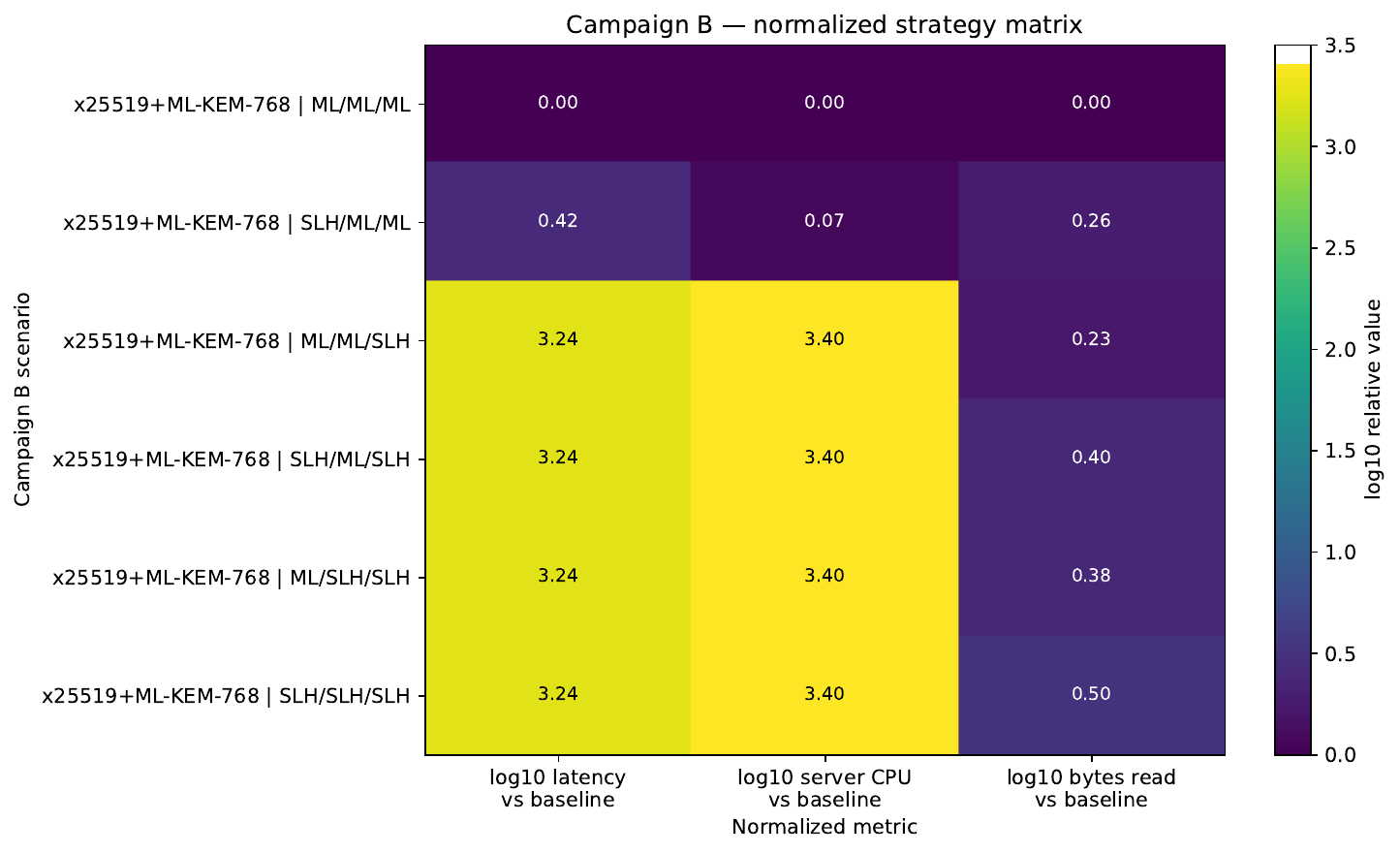}
	\caption{Campaign B: normalized strategy matrix.}
	\label{fig:fig04_campaignB_strategy_heatmap}
\end{figure}

\begin{table}[t]
	\centering
	\caption{Campaign B strategy matrix.}
	\label{tab:block7B_campaignB_strategy_matrix_body}
	\resizebox{\textwidth}{!}{%
	\begin{tabular}{llrrrr}
		\toprule
		scenario\_id & slh\_position\_class & elapsed\_mean\_ms & bytes\_read\_mean & server\_task\_clock\_per\_run\_ms & latency\_relative\_to\_baseline \\
		\midrule
		x25519mlkem768\_\_ml\_root\_\_ml\_int\_\_ml\_leaf & no\_slh & 0.809000 & 16008 & 0.562000 & 1.000000 \\
		x25519mlkem768\_\_slh\_root\_\_ml\_int\_\_ml\_leaf & root & 2.133000 & 28947 & 0.667000 & 2.640000 \\
		x25519mlkem768\_\_ml\_root\_\_ml\_int\_\_slh\_leaf & leaf & 1402.486000 & 27015 & 1401.169000 & 1733.490000 \\
		x25519mlkem768\_\_slh\_root\_\_ml\_int\_\_slh\_leaf & root\_and\_leaf & 1403.166000 & 39954 & 1401.849000 & 1734.330000 \\
		x25519mlkem768\_\_ml\_root\_\_slh\_int\_\_slh\_leaf & intermediate\_and\_leaf & 1405.803000 & 38046 & 1403.109000 & 1737.590000 \\
		x25519mlkem768\_\_slh\_root\_\_slh\_int\_\_slh\_leaf & root\_and\_intermediate\_and\_leaf & 1408.703000 & 50985 & 1404.903000 & 1741.180000 \\
		\bottomrule
	\end{tabular}%
}

\vspace{0.5em}

\resizebox{\textwidth}{!}{%
	\begin{tabular}{llrrrl}
		\toprule
		scenario\_id & slh\_position\_class & server\_cpu\_relative\_to\_baseline & bytes\_read\_relative\_to\_baseline & operational\_plausibility \\
		\midrule
		x25519mlkem768\_\_ml\_root\_\_ml\_int\_\_ml\_leaf & no\_slh & 1.000000 & 1.000000 & Reasonable \\
		x25519mlkem768\_\_slh\_root\_\_ml\_int\_\_ml\_leaf & root & 1.190000 & 1.810000 & Penalized but plausible \\
		x25519mlkem768\_\_ml\_root\_\_ml\_int\_\_slh\_leaf & leaf & 2493.640000 & 1.690000 & Unsuitable for interactive TLS front-end \\
		x25519mlkem768\_\_slh\_root\_\_ml\_int\_\_slh\_leaf & root\_and\_leaf & 2494.850000 & 2.500000 & Unsuitable for interactive TLS front-end \\
		x25519mlkem768\_\_ml\_root\_\_slh\_int\_\_slh\_leaf & intermediate\_and\_leaf & 2497.090000 & 2.380000 & Unsuitable for interactive TLS front-end \\
		x25519mlkem768\_\_slh\_root\_\_slh\_int\_\_slh\_leaf & root\_and\_intermediate\_and\_leaf & 2500.280000 & 3.180000 & Unsuitable for interactive TLS front-end \\
		\bottomrule
	\end{tabular}%
}

\end{table}

\begin{table}[t]
	\centering
	\small
	\setlength{\tabcolsep}{4pt}
	\renewcommand{\arraystretch}{1.08}
	\caption{Campaign B operational plausibility ranking.}
	\label{tab:block7B_campaignB_operational_plausibility_ranking}
	\resizebox{\textwidth}{!}{%
		\begin{tabular}{rlllrrl}
\toprule
plausibility\_rank & scenario\_id & hierarchy\_family\_label & slh\_position\_class & elapsed\_mean\_ms & server\_task\_clock\_per\_run\_ms & operational\_plausibility \\
\midrule
1 & x25519mlkem768\_\_ml\_root\_\_ml\_int\_\_ml\_leaf & ML root / ML int / ML leaf & no\_slh & 0.809000 & 0.562000 & Reasonable \\
2 & x25519mlkem768\_\_slh\_root\_\_ml\_int\_\_ml\_leaf & SLH root / ML int / ML leaf & root & 2.133000 & 0.667000 & Penalized but plausible \\
4 & x25519mlkem768\_\_ml\_root\_\_ml\_int\_\_slh\_leaf & ML root / ML int / SLH leaf & leaf & 1402.486000 & 1401.169000 & Unsuitable for interactive TLS front-end \\
4 & x25519mlkem768\_\_slh\_root\_\_ml\_int\_\_slh\_leaf & SLH root / ML int / SLH leaf & root\_and\_leaf & 1403.166000 & 1401.849000 & Unsuitable for interactive TLS front-end \\
4 & x25519mlkem768\_\_ml\_root\_\_slh\_int\_\_slh\_leaf & ML root / SLH int / SLH leaf & intermediate\_and\_leaf & 1405.803000 & 1403.109000 & Unsuitable for interactive TLS front-end \\
4 & x25519mlkem768\_\_slh\_root\_\_slh\_int\_\_slh\_leaf & SLH root / SLH int / SLH leaf & root\_and\_intermediate\_and\_leaf & 1408.703000 & 1404.903000 & Unsuitable for interactive TLS front-end \\
\bottomrule
\end{tabular}

	}
\end{table}

\begin{figure}[t]
	\centering
	\includegraphics[width=\linewidth]{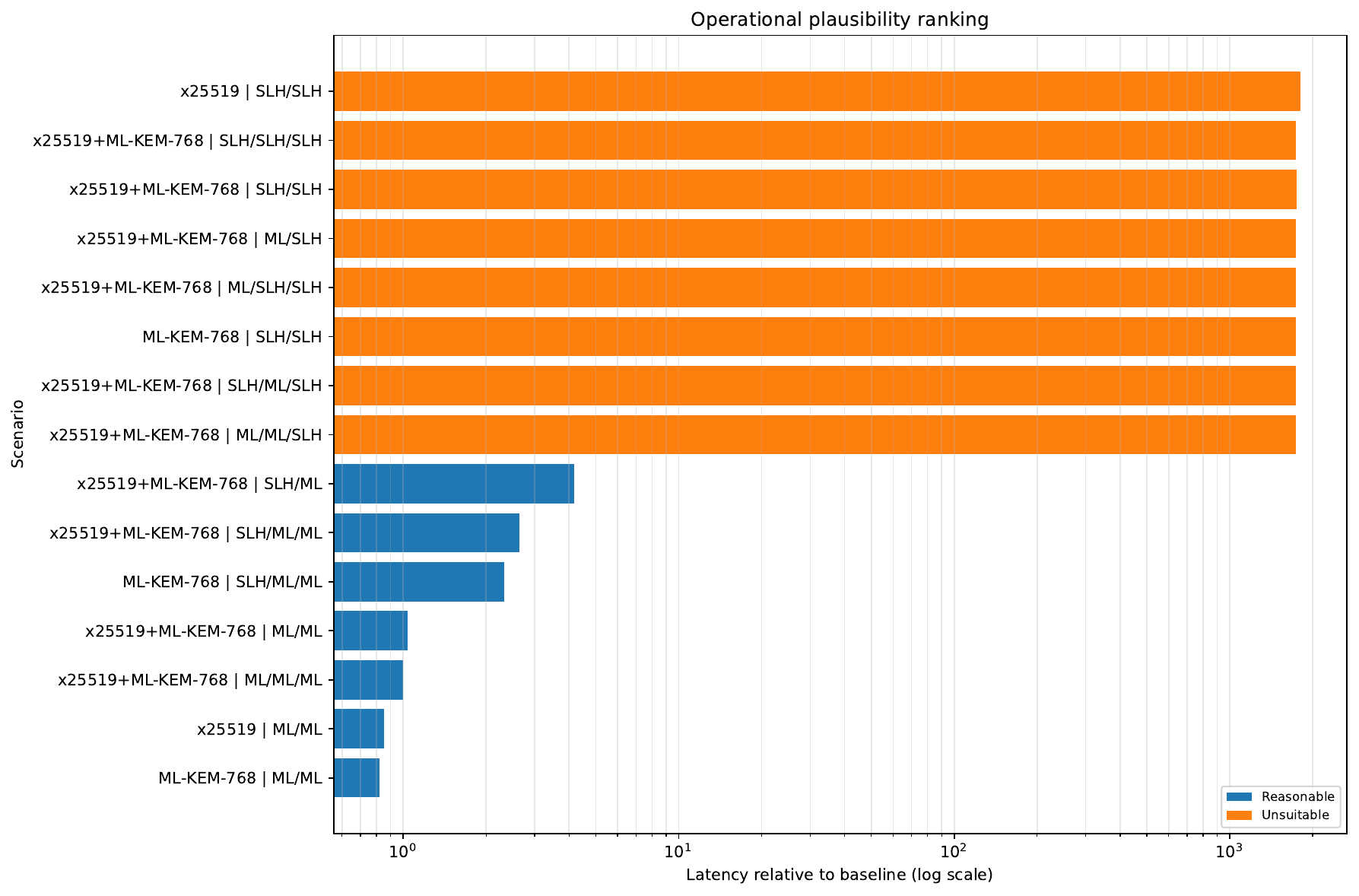}
	\caption{Operational plausibility ranking by scenario, expressed as latency relative to the fully-ML baseline.}
	\label{fig:fig12_operational_plausibility_ranking}
\end{figure}

\subsubsection{Fully-ML baseline}

The baseline strategy for Campaign~B is the hybrid depth-3 hierarchy \\
\texttt{x25519mlkem768\_\_ml\_root\_\_ml\_int\_\_ml\_leaf}. Its mean latency is 0.809~ms, its p95 is 1.000~ms, it reads 16008 bytes per handshake on average, and its server task-clock per run is approximately 0.562~ms. This scenario is the reference point for the relative comparisons reported throughout the paper.

\subsubsection{Root-SLH with ML leaf}

The most important bounded-penalty case in Campaign~B is \\
\texttt{x25519mlkem768\_\_slh\_root\_\_ml\_int\_\_ml\_leaf}. Relative to the fully-ML baseline, its mean latency rises by approximately 2.64$\times$, its server task-clock increases by about 1.19$\times$, and its read volume increases by about 1.81$\times$. The scenario remains within an interactive regime and is the only non-baseline strategy in Campaign~B that remains defensible under a strict operational reading.

\subsubsection{Leaf-SLH strategies}

The remaining mixed strategies in Campaign~B all place SLH-DSA in the leaf. Once that occurs, the design space contracts into a narrow high-cost band. Mean latency ranges from approximately 1402.486~ms to 1408.703~ms, corresponding to roughly 1733.49$\times$ to 1741.18$\times$ the fully-ML baseline. Over the same range, server task-clock grows by approximately 2493.64$\times$ to 2500.28$\times$, while read volume grows by only about 1.69$\times$ to 3.18$\times$.

Once SLH-DSA reaches the leaf, the finer details of the mixed hierarchy explain little of the remaining latency spread. The main explanatory break is between non-leaf-SLH and leaf-SLH configurations, not between individual mixed hierarchies.

\subsubsection{Main result of Campaign~B}

Campaign~B exposes the main strategic pattern of the study. The variable with the greatest explanatory power is not whether SLH-DSA appears somewhere in the hierarchy, but whether it appears in the leaf exposed to the live TLS handshake. Figure~\ref{fig:fig12_operational_plausibility_ranking} shows the pattern compactly: the fully-ML baseline remains in the reasonable region, the \texttt{root SLH / intermediate ML / leaf ML} strategy remains penalized but plausible, and the remaining strategies move into the operationally unsuitable region once SLH-DSA reaches the leaf.

\subsection{Campaign C: depth comparison}

Campaign~C separates topology from signature placement. This campaign shows that increasing hierarchy depth does not behave as a simple monotonic penalty. The cost of moving from depth~2 to depth~3 depends on which certificates become part of the effective chain observed during the handshake.

\begin{figure}[t]
	\centering
	\includegraphics[width=0.95\linewidth]{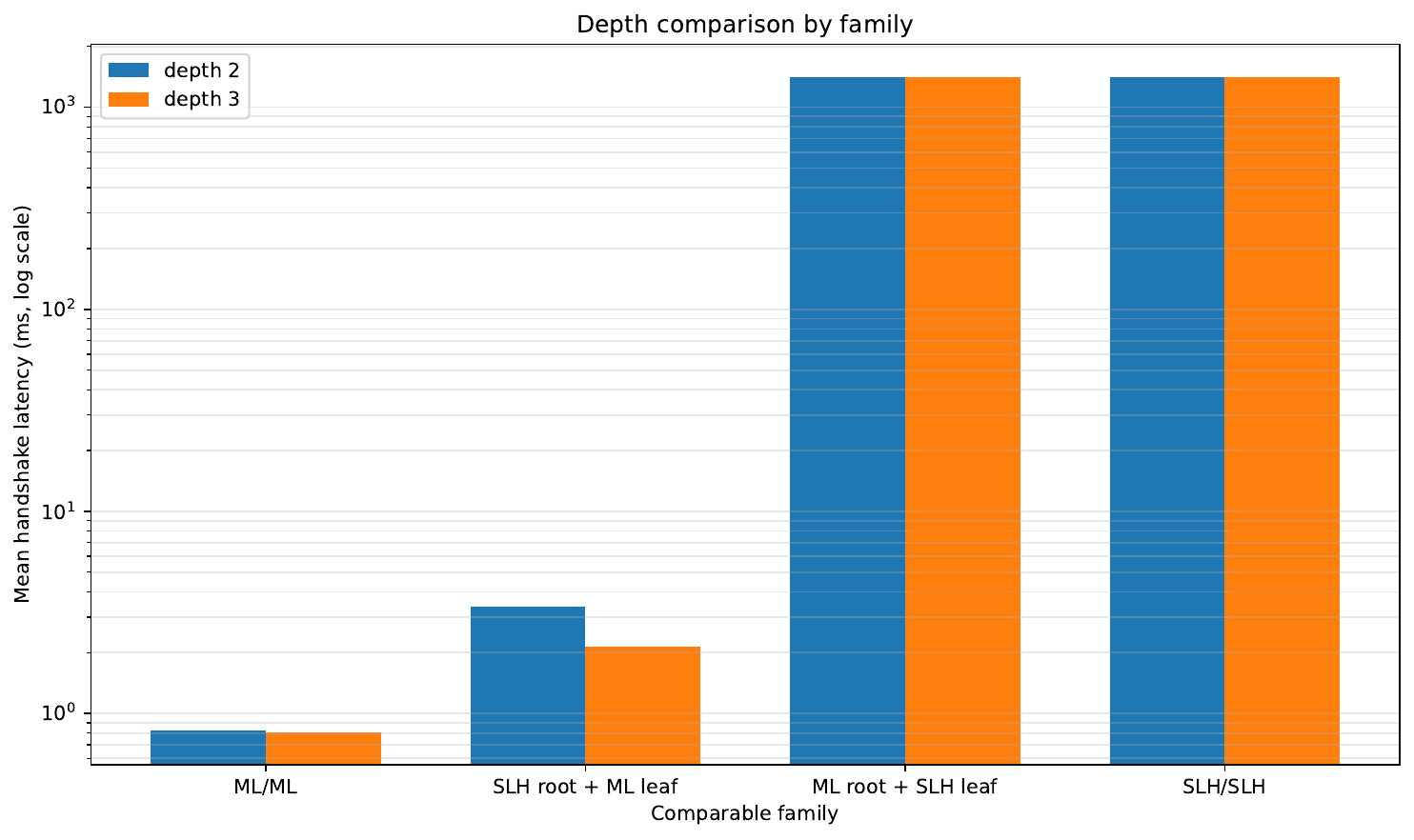}
	\caption{Depth comparison by family.}
	\label{fig:fig05_depth_comparison_by_family}
\end{figure}

\begin{figure}[t]
	\centering
	\includegraphics[width=0.9\linewidth]{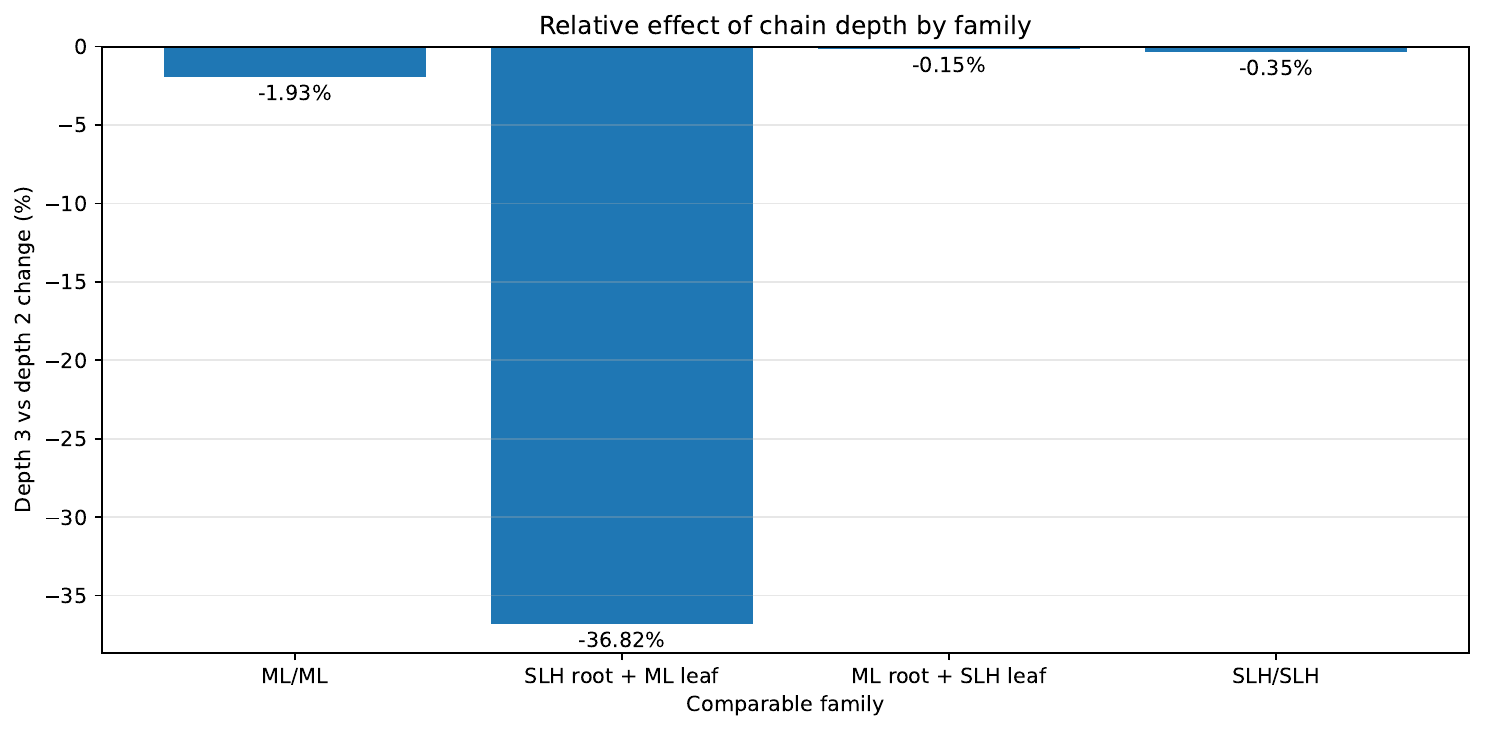}
	\caption{Relative effect of moving from depth~2 to depth~3 by comparable family, expressed as percentage change in mean latency.}
	\label{fig:fig05_depth_comparison_by_family_delta_pct}
\end{figure}

\begin{table}[t]
	\centering
	\small
	\setlength{\tabcolsep}{4pt}
	\renewcommand{\arraystretch}{1.08}
	\caption{Depth~2 vs.\ depth~3 comparisons.}
	\label{tab:block7C_campaignC_depth_comparison_body}
	\resizebox{\textwidth}{!}{%
	\begin{tabular}{L{2.7cm} L{3.5cm} L{3.8cm} r r r r}
		\toprule
		\makecell[l]{pair\\label} &
		\makecell[l]{depth 2\\scenario\_id} &
		\makecell[l]{depth 3\\scenario\_id} &
		\makecell[r]{depth 2\\elapsed mean (ms)} &
		\makecell[r]{depth 3\\elapsed mean (ms)} &
		\makecell[r]{delta elapsed mean (ms)\\d3 minus d2} &
		\makecell[r]{latency ratio\\d3 / d2} \\
		\midrule
		ML/ML &
		\seqsplit{x25519mlkem768\_\_ml\_root\_\_ml\_leaf} &
		\seqsplit{x25519mlkem768\_\_ml\_root\_\_ml\_int\_\_ml\_leaf} &
		0.809300 & 0.809100 & -0.000200 & 0.999700 \\
		SLH root + ML leaf &
		\seqsplit{x25519mlkem768\_\_slh\_root\_\_ml\_leaf} &
		\seqsplit{x25519mlkem768\_\_slh\_root\_\_ml\_int\_\_ml\_leaf} &
		3.376200 & 2.133000 & -1.243100 & 0.631800 \\
		ML root + SLH leaf (ML intermediate) &
		\seqsplit{x25519mlkem768\_\_ml\_root\_\_slh\_leaf} &
		\seqsplit{x25519mlkem768\_\_ml\_root\_\_ml\_int\_\_slh\_leaf} &
		1406.283200 & 1402.486200 & -3.797000 & 0.997300 \\
		ML root + SLH leaf (SLH intermediate) &
		\seqsplit{x25519mlkem768\_\_ml\_root\_\_slh\_leaf} &
		\seqsplit{x25519mlkem768\_\_ml\_root\_\_slh\_int\_\_slh\_leaf} &
		1406.283200 & 1405.802800 & -0.480400 & 0.999700 \\
		SLH/SLH (ML intermediate) &
		\seqsplit{x25519mlkem768\_\_slh\_root\_\_slh\_leaf} &
		\seqsplit{x25519mlkem768\_\_slh\_root\_\_ml\_int\_\_slh\_leaf} &
		1407.714100 & 1403.165900 & -4.548200 & 0.996800 \\
		SLH/SLH (SLH intermediate) &
		\seqsplit{x25519mlkem768\_\_slh\_root\_\_slh\_leaf} &
		\seqsplit{x25519mlkem768\_\_slh\_root\_\_slh\_int\_\_slh\_leaf} &
		1407.714100 & 1408.703300 & 0.989200 & 1.000700 \\
		\bottomrule
	\end{tabular}%
}

\vspace{0.5em}

\resizebox{\textwidth}{!}{%
	\begin{tabular}{L{2.7cm} L{3.5cm} L{3.8cm} r r r r}
		\toprule
		\makecell[l]{pair\\label} &
		\makecell[l]{depth 2\\scenario\_id} &
		\makecell[l]{depth 3\\scenario\_id} &
		\makecell[r]{delta bytes read mean\\d3 minus d2} &
		\makecell[r]{delta chain bytes unique\\d3 minus d2} &
		\makecell[r]{delta server task clock per run (ms)\\d3 minus d2} &
		\makecell[r]{server taskclock ratio\\d3 / d2} \\
		\midrule
		ML/ML &
		\seqsplit{x25519mlkem768\_\_ml\_root\_\_ml\_leaf} &
		\seqsplit{x25519mlkem768\_\_ml\_root\_\_ml\_int\_\_ml\_leaf} &
		16 & 16 & -0.040300 & 0.933100 \\
		SLH root + ML leaf &
		\seqsplit{x25519mlkem768\_\_slh\_root\_\_ml\_leaf} &
		\seqsplit{x25519mlkem768\_\_slh\_root\_\_ml\_int\_\_ml\_leaf} &
		-11015 & -10993 & -1.247500 & 0.348500 \\
		ML root + SLH leaf (ML intermediate) &
		\seqsplit{x25519mlkem768\_\_ml\_root\_\_slh\_leaf} &
		\seqsplit{x25519mlkem768\_\_ml\_root\_\_ml\_int\_\_slh\_leaf} &
		16 & 16 & -3.757500 & 0.997300 \\
		ML root + SLH leaf (SLH intermediate) &
		\seqsplit{x25519mlkem768\_\_ml\_root\_\_slh\_leaf} &
		\seqsplit{x25519mlkem768\_\_ml\_root\_\_slh\_int\_\_slh\_leaf} &
		11047 & 11025 & -1.817400 & 0.998700 \\
		SLH/SLH (ML intermediate) &
		\seqsplit{x25519mlkem768\_\_slh\_root\_\_slh\_leaf} &
		\seqsplit{x25519mlkem768\_\_slh\_root\_\_ml\_int\_\_slh\_leaf} &
		-11015 & -10993 & -3.238100 & 0.997700 \\
		SLH/SLH (SLH intermediate) &
		\seqsplit{x25519mlkem768\_\_slh\_root\_\_slh\_leaf} &
		\seqsplit{x25519mlkem768\_\_slh\_root\_\_slh\_int\_\_slh\_leaf} &
		16 & 16 & -0.183900 & 0.999900 \\
		\bottomrule
	\end{tabular}%
}
\end{table}

\subsubsection{All-ML families}

In the fully-ML family, additional depth does not introduce a meaningful penalty. Comparing
\texttt{x25519mlkem768\_\_ml\_root\_\_ml\_leaf} with
\texttt{x25519mlkem768\_\_ml\_root\_\_ml\_int\_\_ml\_leaf},
the latency ratio is effectively 1.00, with only negligible differences in observed bytes and a slightly smaller server task-clock in the depth-3 case. Depth alone is therefore not enough to degrade a well-behaved ML hierarchy in any meaningful way.

\subsubsection{Root-SLH with ML leaf}

The clearest topological effect appears when the root uses SLH-DSA while the leaf remains ML-DSA. In that family, the depth-3 variant is substantially cheaper than the depth-2 one. The latency ratio falls to approximately 0.6318, corresponding to a reduction of about 36.8\% in mean latency. At the same time, \texttt{bytes\_read\_mean} decreases by about 11015 bytes and \texttt{chain\_bytes\_unique} by about 10993 bytes.

The interpretation follows from effective chain exposure. In the depth-2 variant, the observed chain corresponds to root plus leaf; in the depth-3 variant, it corresponds to intermediate plus leaf. Since the root is the heavy SLH signer in this family, removing it from the observed chain reduces visible transport and lowers overall cost materially.

\subsubsection{Leaf-SLH families}

The leaf-SLH families show the opposite pattern. Once SLH-DSA reaches the leaf, changes in depth still affect transport, sometimes substantially, but they have little effect on latency. In the \texttt{ML root + SLH leaf} family, the depth ratio remains effectively 1.00 even when one depth-3 variant increases observed transport by more than 11~KB. In the \texttt{SLH/SLH} family, depth may either reduce or increase observed transport, yet latency remains nearly unchanged.

This pattern indicates that depth continues to matter for transport while losing explanatory power over the main latency regime once the leaf itself becomes the dominant source of cryptographic cost.

\subsubsection{Main result of Campaign~C}

Campaign~C shows that hierarchy depth is not, by itself, a cost verdict. Whether additional depth penalizes, preserves, or reduces observed cost depends on how the topology changes the effective chain seen during the handshake. This matters for the interpretation of post-quantum certificate hierarchies because logical depth and practical exposure cannot be treated as interchangeable notions.

\subsection{Campaign D: KEX mode exploration}

Campaign~D tests whether moving from classical to hybrid, or from hybrid to pure post-quantum key establishment, materially changes the placement-driven picture once comparable chains are held fixed.

\begin{figure}[t]
	\centering
	\includegraphics[width=\linewidth]{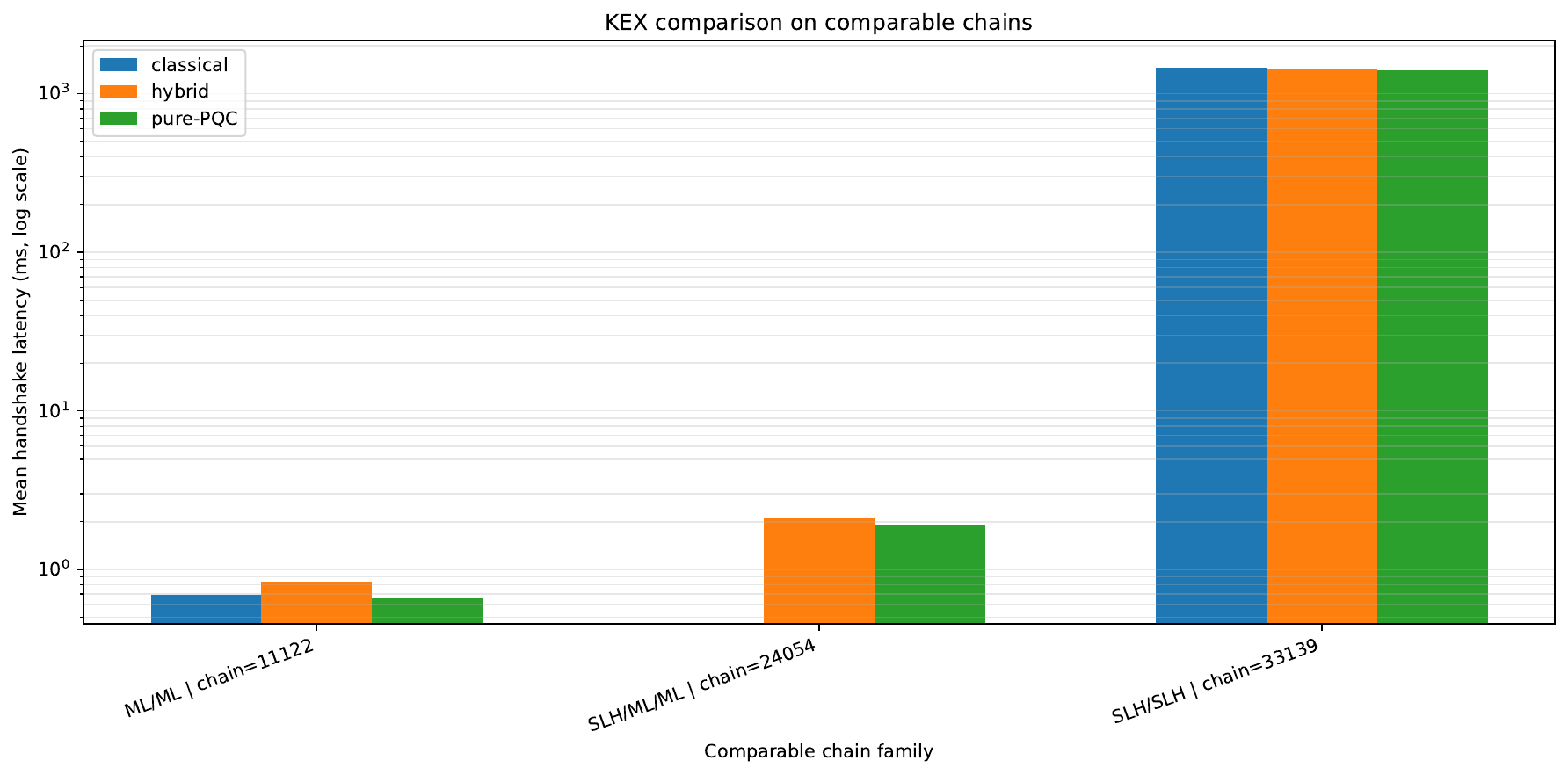}
	\caption{KEX comparison on comparable chains.}
	\label{fig:fig13_kex_comparison_comparable_chains}
\end{figure}

\begin{table}[t]
	\centering
	\caption{KEX comparisons on comparable chains.}
	\label{tab:block7D_campaignD_kex_comparisons_body}
	\resizebox{\textwidth}{!}{%
	\begin{tabular}{lllllrrrr}
		\toprule
		comparison\_type & family\_label & from\_kex\_mode & to\_kex\_mode & leaf\_family & depth & elapsed\_mean\_from\_ms & elapsed\_mean\_to\_ms & latency\_ratio\_to\_over\_from \\
		\midrule
		classical\_vs\_hybrid & ML root / ML leaf (depth 2) & classical & hybrid & ML-DSA & 2 & 0.6885 & 0.8406 & 1.2210 \\
		classical\_vs\_hybrid & SLH root / SLH leaf (depth 2) & classical & hybrid & SLH-DSA & 2 & 1464.9327 & 1413.9914 & 0.9652 \\
		hybrid\_vs\_pure\_pqc & ML root / ML leaf (depth 2) & hybrid & pure\_pqc & ML-DSA & 2 & 0.8093 & 0.6652 & 0.8220 \\
		hybrid\_vs\_pure\_pqc & SLH root / ML int / ML leaf (depth 3) & hybrid & pure\_pqc & ML-DSA & 3 & 2.1330 & 1.8842 & 0.8833 \\
		hybrid\_vs\_pure\_pqc & SLH root / SLH leaf (depth 2) & hybrid & pure\_pqc & SLH-DSA & 2 & 1407.7141 & 1405.4559 & 0.9984 \\
		\bottomrule
	\end{tabular}%
}

\vspace{0.5em}

\resizebox{\textwidth}{!}{%
	\begin{tabular}{lllllrrrr}
		\toprule
		comparison\_type & family\_label & from\_kex\_mode & to\_kex\_mode & leaf\_family & depth & bytes\_read\_from & bytes\_read\_to & bytes\_read\_ratio\_to\_over\_from \\
		\midrule
		classical\_vs\_hybrid & ML root / ML leaf (depth 2) & classical & hybrid & ML-DSA & 2 & 14904 & 15992 & 1.0730 \\
		classical\_vs\_hybrid & SLH root / SLH leaf (depth 2) & classical & hybrid & SLH-DSA & 2 & 49881 & 50969 & 1.0218 \\
		hybrid\_vs\_pure\_pqc & ML root / ML leaf (depth 2) & hybrid & pure\_pqc & ML-DSA & 2 & 15992 & 15960 & 0.9980 \\
		hybrid\_vs\_pure\_pqc & SLH root / ML int / ML leaf (depth 3) & hybrid & pure\_pqc & ML-DSA & 3 & 28947 & 28915 & 0.9989 \\
		hybrid\_vs\_pure\_pqc & SLH root / SLH leaf (depth 2) & hybrid & pure\_pqc & SLH-DSA & 2 & 50969 & 50937 & 0.9994 \\
		\bottomrule
	\end{tabular}%
}

\vspace{0.5em}

\resizebox{\textwidth}{!}{%
	\begin{tabular}{lllllrrrr}
		\toprule
		comparison\_type & family\_label & from\_kex\_mode & to\_kex\_mode & leaf\_family & depth & server\_task\_from\_ms & server\_task\_to\_ms & server\_task\_ratio\_to\_over\_from \\
		\midrule
		classical\_vs\_hybrid & ML root / ML leaf (depth 2) & classical & hybrid & ML-DSA & 2 & 0.5287 & 0.6228 & 1.1780 \\
		classical\_vs\_hybrid & SLH root / SLH leaf (depth 2) & classical & hybrid & SLH-DSA & 2 & 1462.2813 & 1411.3448 & 0.9652 \\
		hybrid\_vs\_pure\_pqc & ML root / ML leaf (depth 2) & hybrid & pure\_pqc & ML-DSA & 2 & 0.6022 & 0.5152 & 0.8555 \\
		hybrid\_vs\_pure\_pqc & SLH root / ML int / ML leaf (depth 3) & hybrid & pure\_pqc & ML-DSA & 3 & 0.6672 & 0.5273 & 0.7903 \\
		hybrid\_vs\_pure\_pqc & SLH root / SLH leaf (depth 2) & hybrid & pure\_pqc & SLH-DSA & 2 & 1405.0874 & 1402.8960 & 0.9984 \\
		\bottomrule
	\end{tabular}%
}

\end{table}

In ML-leaf regimes, KEX choice modulates the cost profile. For example, moving from classical to hybrid in the simple ML-leaf case increases latency by about 1.2210$\times$, while moving from hybrid to pure post-quantum reduces it to about 0.8220$\times$ the hybrid value. Similar but bounded effects appear in the \texttt{SLH root / ML int / ML leaf} family, where hybrid to pure post-quantum produces a latency ratio of about 0.8833$\times$.

Once the leaf uses SLH-DSA, KEX mode contributes only local variation within the dominant heavy regime. In the \texttt{SLH root / SLH leaf} comparison, classical to hybrid yields a latency ratio of about 0.9652$\times$, and hybrid to pure post-quantum about 0.9984$\times$. These ratios are local perturbations inside an already severe regime.

\subsubsection{Main result of Campaign~D}

Campaign~D shows that KEX mode matters in low-cost ML-leaf scenarios and can produce moderate changes in bounded regimes. Pure post-quantum key establishment does not change the main picture. The decisive variable remains signature placement, especially the decision to place SLH-DSA in the handshake-exposed server leaf.

\section{Cross-Cutting Cryptographic Interpretation}

The campaign-based results establish the empirical structure of the study. Reorganizing the evidence around four cross-cutting questions clarifies that structure: whether signature placement explains behavior better than signature-family presence alone, whether transport-related overhead is sufficient to explain the heavy regime, whether effective chain exposure must be analyzed as part of the authentication path, and whether the leaf-SLH cases form a distinct operational regime or the upper tail of a continuous cost spectrum. Under this organization, the results support an interpretation of post-quantum TLS authentication as a hierarchy-sensitive cryptographic design problem.

\subsection{Signature placement versus mere signature presence}

\begin{figure}[t]
	\centering
	\includegraphics[width=0.9\linewidth]{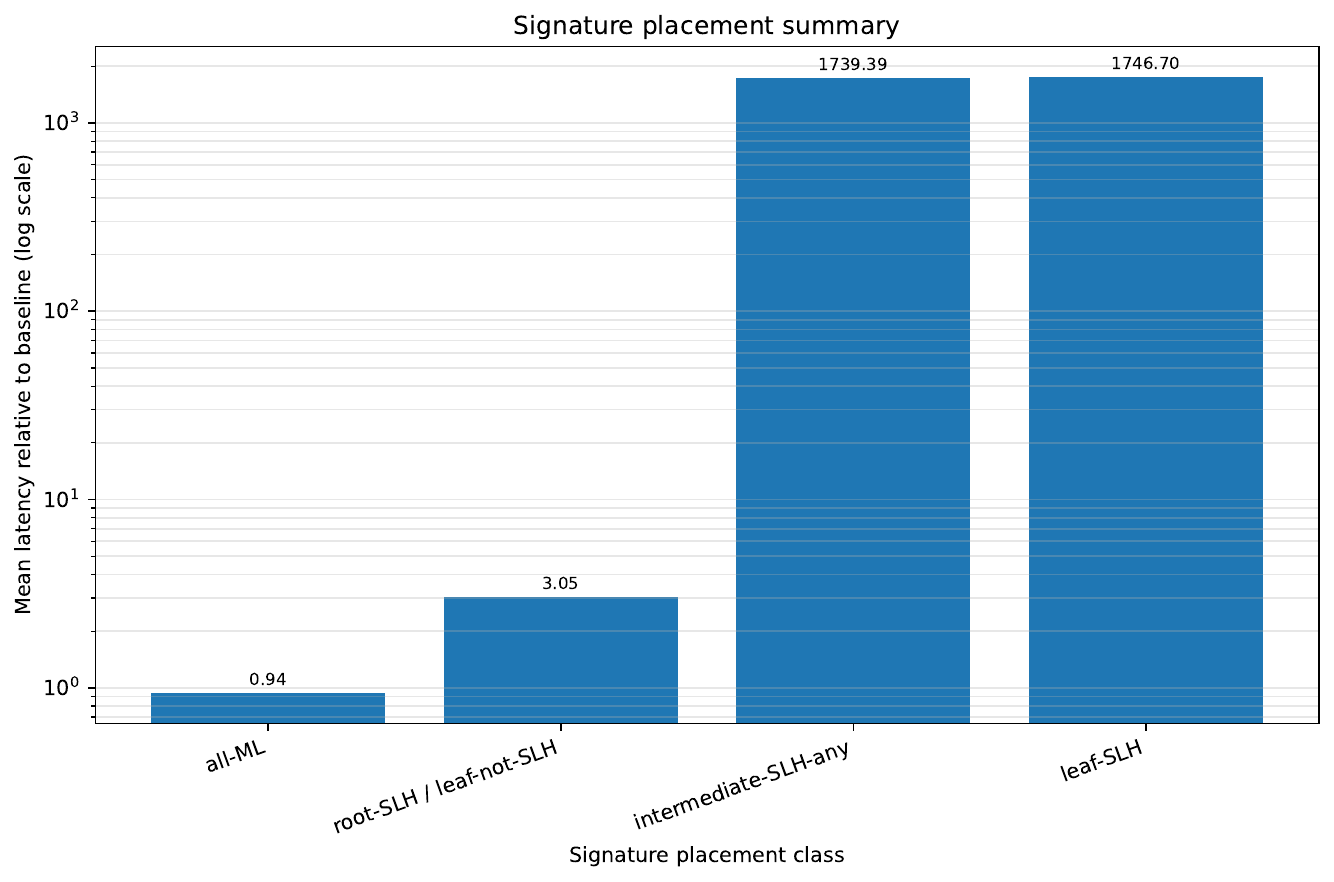}
	\caption{Signature placement summary.}
	\label{fig:fig11_signature_placement_summary}
\end{figure}

\begin{table}[t]
	\centering
	\caption{Signature placement summary.}
	\label{tab:block8_signature_placement_summary_body}
	\resizebox{\textwidth}{!}{%
	\begin{tabular}{lrrrrrr}
		\toprule
		placement\_class & n\_scenarios & mean\_elapsed\_ms & median\_elapsed\_ms & min\_elapsed\_ms & max\_elapsed\_ms & mean\_latency\_vs\_baseline \\
		\midrule
		all\_ml & 5 & 0.763 & 0.809 & 0.665 & 0.841 & 0.942 \\
		root\_slh\_leaf\_not\_slh & 3 & 2.464 & 2.133 & 1.884 & 3.376 & 3.046 \\
		intermediate\_slh\_any & 2 & 1407.253 & 1407.253 & 1405.803 & 1408.703 & 1739.386 \\
		leaf\_slh & 9 & 1413.171 & 1406.283 & 1402.486 & 1464.933 & 1746.701 \\
		\bottomrule
	\end{tabular}%
}

\vspace{0.5em}

\resizebox{\textwidth}{!}{%
	\begin{tabular}{lrrrrrr}
		\toprule
		placement\_class & median\_latency\_vs\_baseline & mean\_bytes\_vs\_baseline & mean\_server\_cpu\_vs\_baseline & mean\_server\_over\_elapsed & mean\_client\_over\_elapsed \\
		\midrule
		all\_ml & 1.000 & 0.985 & 1.008 & 0.744 & 0.742 \\
		root\_slh\_leaf\_not\_slh & 2.636 & 2.037 & 1.844 & 0.387 & 0.803 \\
		intermediate\_slh\_any & 1739.386 & 2.781 & 2498.686 & 0.998 & 0.003 \\
		leaf\_slh & 1738.188 & 2.678 & 2510.849 & 0.998 & 0.002 \\
		\bottomrule
	\end{tabular}%
}
\end{table}

The central empirical finding is that signature placement explains the observed authentication regimes better than signature-family presence alone. When scenarios are grouped by placement class rather than by campaign, the structure of the data becomes more clearly separated.

Figure~\ref{fig:fig11_signature_placement_summary} and Table~\ref{tab:block8_signature_placement_summary_body} organize the scenario space into four categories: all-ML, root-SLH with non-SLH leaf, intermediate-SLH-any, and leaf-SLH. The all-ML class defines the low-cost reference regime of the study. Its mean latency is approximately 0.7625~ms, with mean server-side compute essentially aligned with the baseline. These scenarios form a tight and stable low-cost cluster.

The \texttt{root-SLH / leaf-not-SLH} class is clearly more expensive than all-ML, but it remains within a bounded interactive regime. Its mean latency rises to approximately 2.4645~ms, corresponding to a mean multiplier of about 3.0461$\times$ relative to baseline, while server-side compute rises by about 1.8444$\times$. The presence of SLH-DSA somewhere in the hierarchy therefore does not by itself force the handshake into the catastrophic region.

By contrast, the \texttt{leaf-SLH} class defines the heavy regime of the study. Its mean latency reaches approximately 1413.1706~ms, with a mean latency multiplier of approximately 1746.7007$\times$ relative to baseline and a mean server-side compute multiplier of approximately 2510.8488$\times$. This contrast is too large to read as a continuation of the upper-layer SLH penalty. Moving from all-ML to upper-layer SLH produces a bounded increase, whereas moving from upper-layer SLH to leaf-SLH produces a discontinuity.

This is the clearest expression of the paper's main thesis. The decisive variable is not whether SLH-DSA appears somewhere in the certification path, but whether it reaches the handshake-exposed leaf. Hierarchy position therefore has greater explanatory force than signature-family presence taken in the abstract.

\subsection{Transport-related overhead versus cryptographic cost}

\begin{figure}[t]
	\centering
	\includegraphics[width=0.9\linewidth]{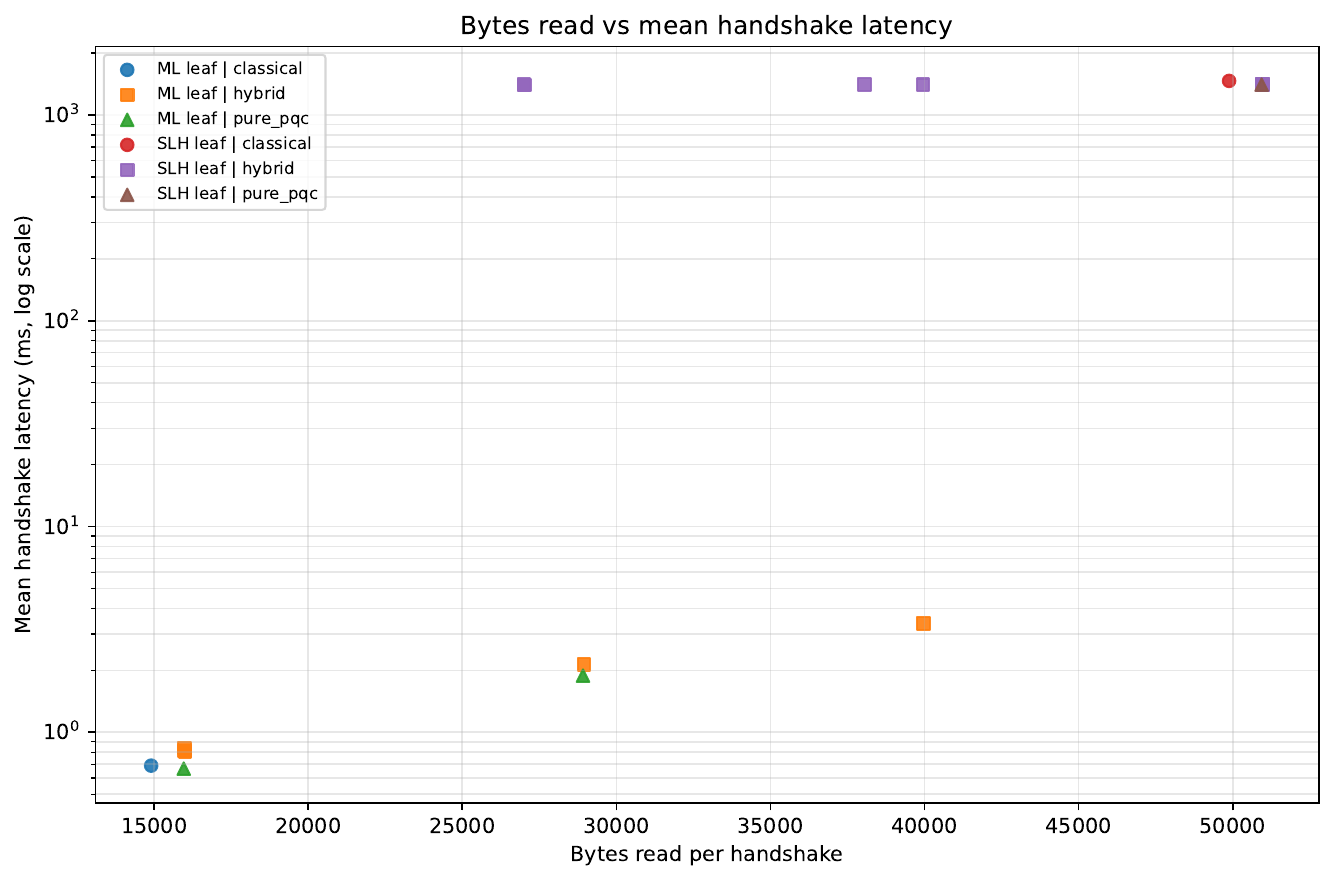}
	\caption{Bytes read versus mean handshake latency.}
	\label{fig:fig06_bytes_read_vs_mean_latency}
\end{figure}

\begin{figure}[t]
	\centering
	\includegraphics[width=0.9\linewidth]{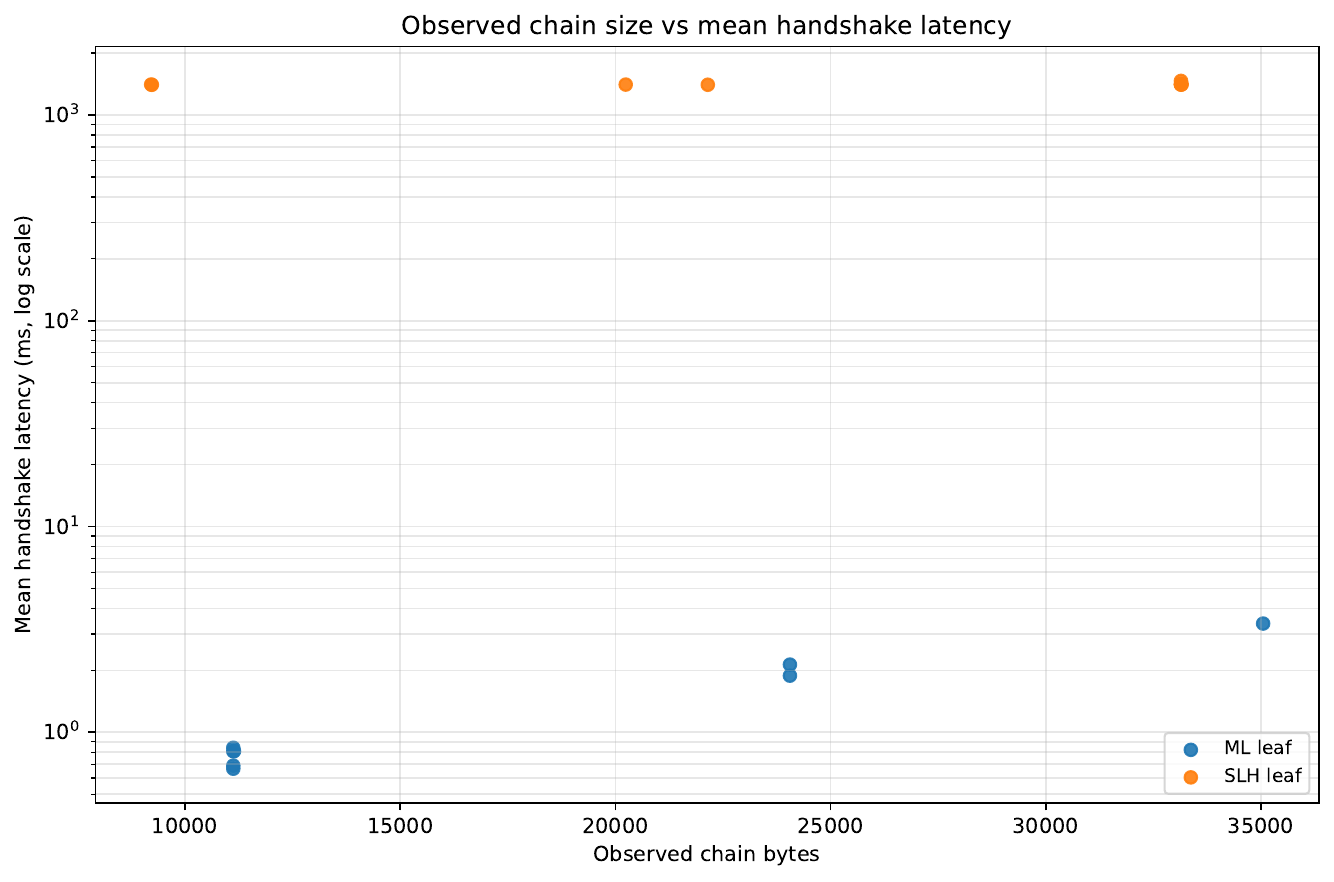}
	\caption{Observed chain size versus mean handshake latency.}
	\label{fig:fig07_chain_bytes_vs_mean_latency}
\end{figure}

\begin{figure}[t]
	\centering
	\includegraphics[width=\linewidth]{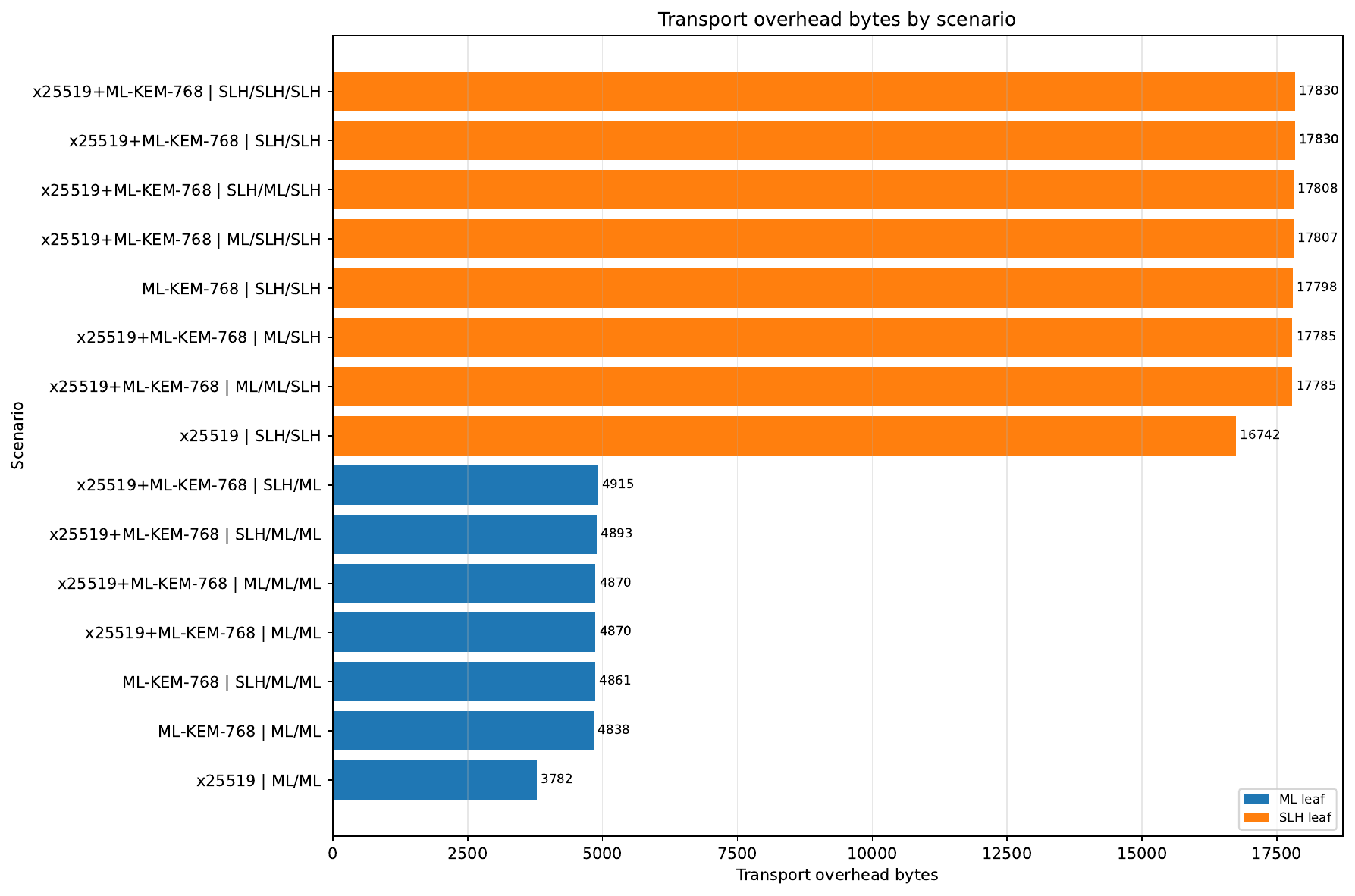}
	\caption{Transport overhead bytes by scenario.}
	\label{fig:fig18_transport_overhead_bytes}
\end{figure}

\begin{table}[t]
	\centering
	\caption{Transport-versus-latency correlations.}
	\label{tab:block9_transport_vs_crypto_correlations_body}
	\begin{tabular}{lrlrr}
\toprule
subset & n\_scenarios & metric & pearson\_r & spearman\_rho \\
\midrule
all\_scenarios & 17 & bytes\_read\_mean & 0.7493 & 0.8503 \\
all\_scenarios & 17 & chain\_bytes\_unique & 0.3943 & 0.5227 \\
non\_leaf\_slh & 8 & bytes\_read\_mean & 0.9937 & 0.8982 \\
non\_leaf\_slh & 8 & chain\_bytes\_unique & 0.9933 & 0.8045 \\
leaf\_slh\_only & 9 & bytes\_read\_mean & 0.3518 & 0.5523 \\
leaf\_slh\_only & 9 & chain\_bytes\_unique & 0.3826 & 0.5919 \\
\bottomrule
\end{tabular}

\end{table}

\begin{table}[t]
	\centering
	\small
	\setlength{\tabcolsep}{4pt}
	\renewcommand{\arraystretch}{1.08}
	\caption{Counterexamples to wire-size-only explanations.}
	\label{tab:block9_transport_vs_crypto_counterexamples_body}
	\resizebox{\textwidth}{!}{%
	\begin{tabular}{r l L{3.6cm} L{3.6cm} r r r r r r}
		\toprule
		rank &
		\makecell[l]{transport\\metric} &
		\makecell[l]{scenario more bytes\\lower latency} &
		\makecell[l]{scenario less bytes\\higher latency} &
		\makecell[r]{more bytes\\value} &
		\makecell[r]{less bytes\\value} &
		\makecell[r]{bytes\\diff} &
		\makecell[r]{lower\\latency (ms)} &
		\makecell[r]{higher\\latency (ms)} &
		\makecell[r]{latency ratio\\higher / lower} \\
		\midrule
		1 & bytes\_read\_mean &
		\seqsplit{x25519mlkem768\_\_slh\_root\_\_ml\_leaf} &
		\seqsplit{x25519mlkem768\_\_ml\_root\_\_slh\_leaf} &
		39962.0000 & 26999.0000 & 12963.0000 & 3.3762 & 1406.2832 & 416.5316 \\
		2 & bytes\_read\_mean &
		\seqsplit{x25519mlkem768\_\_slh\_root\_\_ml\_leaf} &
		\seqsplit{x25519mlkem768\_\_ml\_root\_\_ml\_int\_\_slh\_leaf} &
		39962.0000 & 27015.0000 & 12947.0000 & 3.3762 & 1402.4862 & 415.4070 \\
		3 & bytes\_read\_mean &
		\seqsplit{mlkem768\_\_slh\_root\_\_ml\_int\_\_ml\_leaf} &
		\seqsplit{x25519mlkem768\_\_ml\_root\_\_slh\_leaf} &
		28915.0000 & 26999.0000 & 1916.0000 & 1.8842 & 1406.2832 & 746.3713 \\
		4 & bytes\_read\_mean &
		\seqsplit{x25519mlkem768\_\_slh\_root\_\_ml\_int\_\_ml\_leaf} &
		\seqsplit{x25519mlkem768\_\_ml\_root\_\_slh\_leaf} &
		28947.0000 & 26999.0000 & 1948.0000 & 2.1330 & 1406.2832 & 659.2901 \\
		5 & bytes\_read\_mean &
		\seqsplit{mlkem768\_\_slh\_root\_\_ml\_int\_\_ml\_leaf} &
		\seqsplit{x25519mlkem768\_\_ml\_root\_\_ml\_int\_\_slh\_leaf} &
		28915.0000 & 27015.0000 & 1900.0000 & 1.8842 & 1402.4862 & 744.3561 \\
		\bottomrule
	\end{tabular}%
}

\vspace{0.8em}

\resizebox{\textwidth}{!}{%
	\begin{tabular}{r l L{3.6cm} L{3.6cm} r r r r r r}
		\toprule
		rank &
		\makecell[l]{transport\\metric} &
		\makecell[l]{scenario more bytes\\lower latency} &
		\makecell[l]{scenario less bytes\\higher latency} &
		\makecell[r]{more bytes\\value} &
		\makecell[r]{less bytes\\value} &
		\makecell[r]{bytes\\diff} &
		\makecell[r]{lower\\latency (ms)} &
		\makecell[r]{higher\\latency (ms)} &
		\makecell[r]{latency ratio\\higher / lower} \\
		\midrule
		1 & chain\_bytes\_unique &
		\seqsplit{x25519mlkem768\_\_slh\_root\_\_ml\_leaf} &
		\seqsplit{x25519mlkem768\_\_ml\_root\_\_slh\_leaf} &
		35047.0000 & 9214.0000 & 25833.0000 & 3.3762 & 1406.2832 & 416.5316 \\
		2 & chain\_bytes\_unique &
		\seqsplit{x25519mlkem768\_\_slh\_root\_\_ml\_leaf} &
		\seqsplit{x25519mlkem768\_\_ml\_root\_\_ml\_int\_\_slh\_leaf} &
		35047.0000 & 9230.0000 & 25817.0000 & 3.3762 & 1402.4862 & 415.4070 \\
		3 & chain\_bytes\_unique &
		\seqsplit{mlkem768\_\_slh\_root\_\_ml\_int\_\_ml\_leaf} &
		\seqsplit{x25519mlkem768\_\_ml\_root\_\_slh\_leaf} &
		24054.0000 & 9214.0000 & 14840.0000 & 1.8842 & 1406.2832 & 746.3713 \\
		4 & chain\_bytes\_unique &
		\seqsplit{mlkem768\_\_slh\_root\_\_ml\_int\_\_ml\_leaf} &
		\seqsplit{x25519mlkem768\_\_ml\_root\_\_ml\_int\_\_slh\_leaf} &
		24054.0000 & 9230.0000 & 14824.0000 & 1.8842 & 1402.4862 & 744.3561 \\
		5 & chain\_bytes\_unique &
		\seqsplit{x25519mlkem768\_\_slh\_root\_\_ml\_int\_\_ml\_leaf} &
		\seqsplit{x25519mlkem768\_\_ml\_root\_\_slh\_leaf} &
		24054.0000 & 9214.0000 & 14840.0000 & 2.1330 & 1406.2832 & 659.2901 \\
		\bottomrule
	\end{tabular}%
}
\end{table}

Post-quantum TLS evaluation can easily reduce certificate overhead to wire cost. The data support a more limited interpretation. Transport-related overhead is relevant and, outside the heavy regime, often highly informative. It is not sufficient to explain the catastrophic behavior associated with leaf-SLH scenarios.

Across the full dataset, latency exhibits a moderate positive relationship with \texttt{bytes\_read\_mean} and a weaker one with \texttt{chain\_bytes\_unique}. This could suggest that the heavy scenarios are the consequence of larger transmitted objects. Once leaf-SLH scenarios are excluded, however, latency in the non-catastrophic region tracks both \texttt{bytes\_read\_mean} and \texttt{chain\_bytes\_unique} very closely. Transport explains much of the variation while the system remains in a bounded regime.

The picture changes within the leaf-SLH cases. In that region, the relation between latency and transport weakens sharply. Very large latency persists despite only moderate variation in transferred bytes and observed chain size. Figures~\ref{fig:fig06_bytes_read_vs_mean_latency} and~\ref{fig:fig07_chain_bytes_vs_mean_latency} show this geometrically: outside leaf-SLH, the points follow a relatively coherent growth pattern; inside leaf-SLH, they collapse into a heavy cloud whose internal latency variation is only weakly explained by transport.

The counterexamples in Table~\ref{tab:block9_transport_vs_crypto_counterexamples_body} are especially informative. A scenario may transmit more bytes and still remain hundreds of times faster than another scenario that transmits fewer bytes, provided the former keeps SLH-DSA out of the leaf while the latter places it directly in the interactive certificate. That pattern cannot be reduced to wire expansion alone.

The interpretation is layered. Transport overhead is real and remains highly relevant in bounded regimes. Once SLH-DSA reaches the leaf, however, the dominant explanatory variable changes. At that point, the principal cost source is concentrated cryptographic work during live authentication.

\subsection{Effective chain exposure as a first-class variable}

The transport findings become fully intelligible only when effective chain exposure is treated as a first-class analytical variable. The logical PKI hierarchy is not the whole story. What matters for the handshake is the chain that is actually exposed and processed during protocol execution.

Campaign~C provides the clearest evidence for this point. In the \texttt{root-SLH / leaf-ML} family, moving from depth~2 to depth~3 reduces latency substantially instead of increasing it. In the depth~2 case, the observed chain corresponds to root plus leaf; in the depth~3 case, it corresponds to intermediate plus leaf. Since the root is the heavy SLH signer in that family, removing it from the effective chain reduces visible transport and lowers overall cost materially.

Effective chain exposure is part of the cryptographic-protocol phenomenon under study. It determines which certificates are transmitted, which signature objects enter the authenticated path, and which validation burden becomes visible during the handshake.

The point is especially important in post-quantum settings because certificate families may differ sharply in footprint and processing cost. Under those conditions, the gap between declared hierarchy and effective exposure can become operationally decisive. A study that models only logical hierarchy depth, while ignoring which certificates appear in the live chain, risks missing one of the mechanisms through which authentication cost is shaped.

For that reason, this paper treats effective chain exposure as one of the main explanatory variables. The distinction between root-plus-leaf and intermediate-plus-leaf is one of the mechanisms through which hierarchy design changes the cryptographic surface of the TLS handshake.

\subsection{Emergence of a distinct leaf-SLH regime}

The evidence presented so far supports a stronger claim than the statement that some scenarios are slower than others. It supports the conclusion that the leaf-SLH cases form a distinct operational regime.

This regime is characterized by four simultaneous properties. First, it exhibits a stable latency plateau around 1.4 seconds rather than a broad gradient of increasingly bad cases. Second, it is tightly linked to signature placement: once SLH-DSA reaches the leaf, the heavy regime appears consistently across campaigns, topology variants, and KEX modes. Third, its relation to transport weakens sharply, as shown by the transport correlations and the wire-size counterexamples. Fourth, it is comparatively insensitive to changes in key-establishment mode, as Campaign~D shows.

Taken together, these properties indicate that leaf-SLH is not simply the expensive end of an otherwise continuous design space. It is a qualitatively different authentication regime with its own internal regularity. The scenarios in that class vary in transmitted bytes, chain composition, and hierarchy structure, but they remain confined to a common heavy band whose defining feature is the placement of SLH-DSA in the handshake-exposed leaf.

The paper therefore treats leaf-SLH as a regime concept rather than as a single bad design point. The experimental campaigns reveal a recurring structure in which certificate placement drives the handshake into a state that is only weakly explained by transport, highly stable across scenario variants, and operationally distinct from both the all-ML baseline and the bounded upper-layer-SLH cases.

The next section examines where the cost of this regime resides. If leaf-SLH defines a distinct state of TLS authentication, the next causal question is whether that state is balanced across both peers, validation-skewed, or overwhelmingly concentrated on the server side.

\section{Client/Server Cost Decomposition}

The previous sections established that leaf-SLH scenarios form a distinct heavy regime and that transport-related overhead alone does not explain their cost. This section examines where active cryptographic work is incurred during the handshake by decomposing the authentication path into client- and server-side measurements.

\begin{figure}[t]
	\centering
	\includegraphics[width=\linewidth]{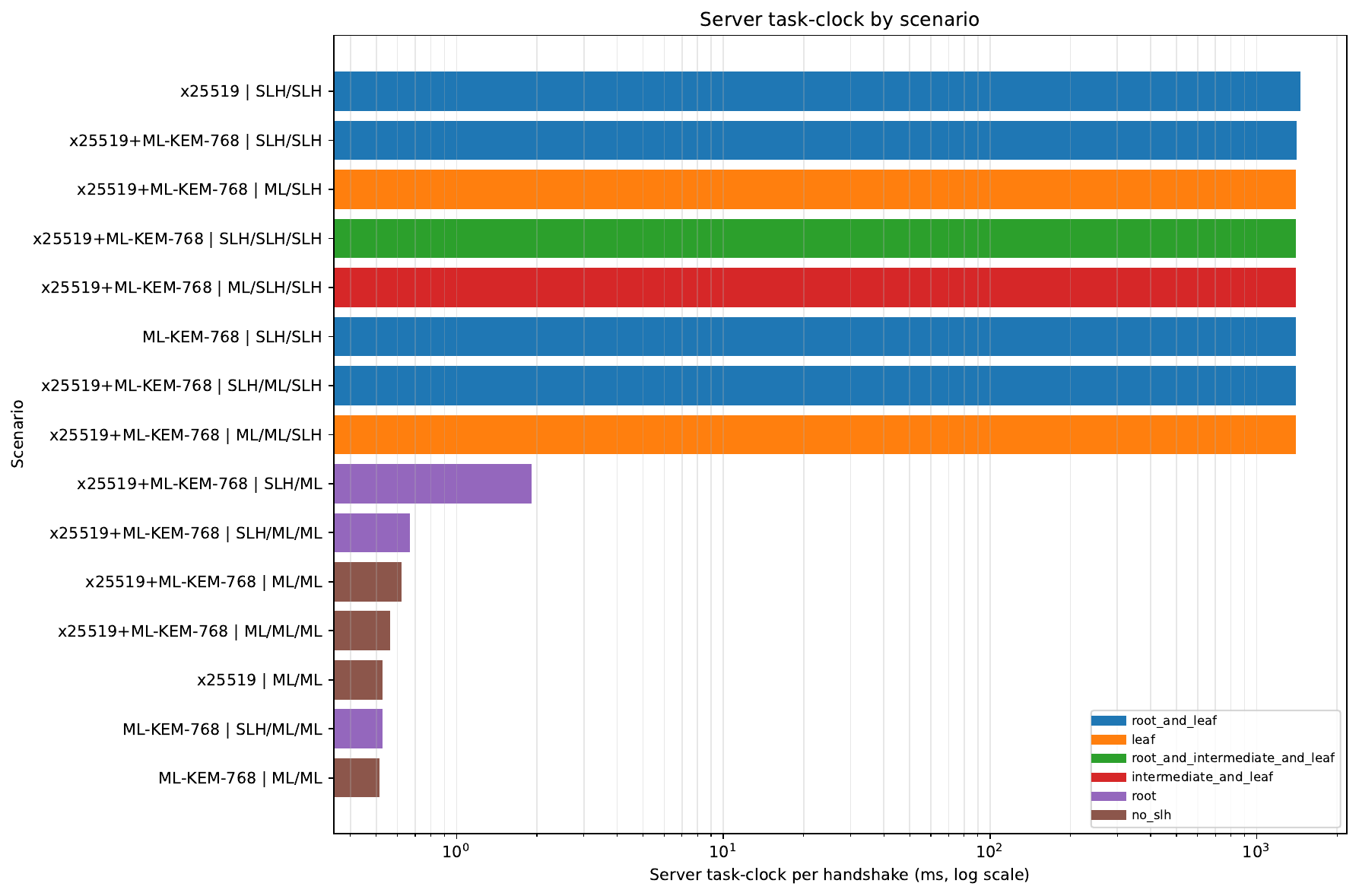}
	\caption{Server task-clock by scenario.}
	\label{fig:fig08_server_taskclock_by_scenario}
\end{figure}

\begin{figure}[t]
	\centering
	\includegraphics[width=\linewidth]{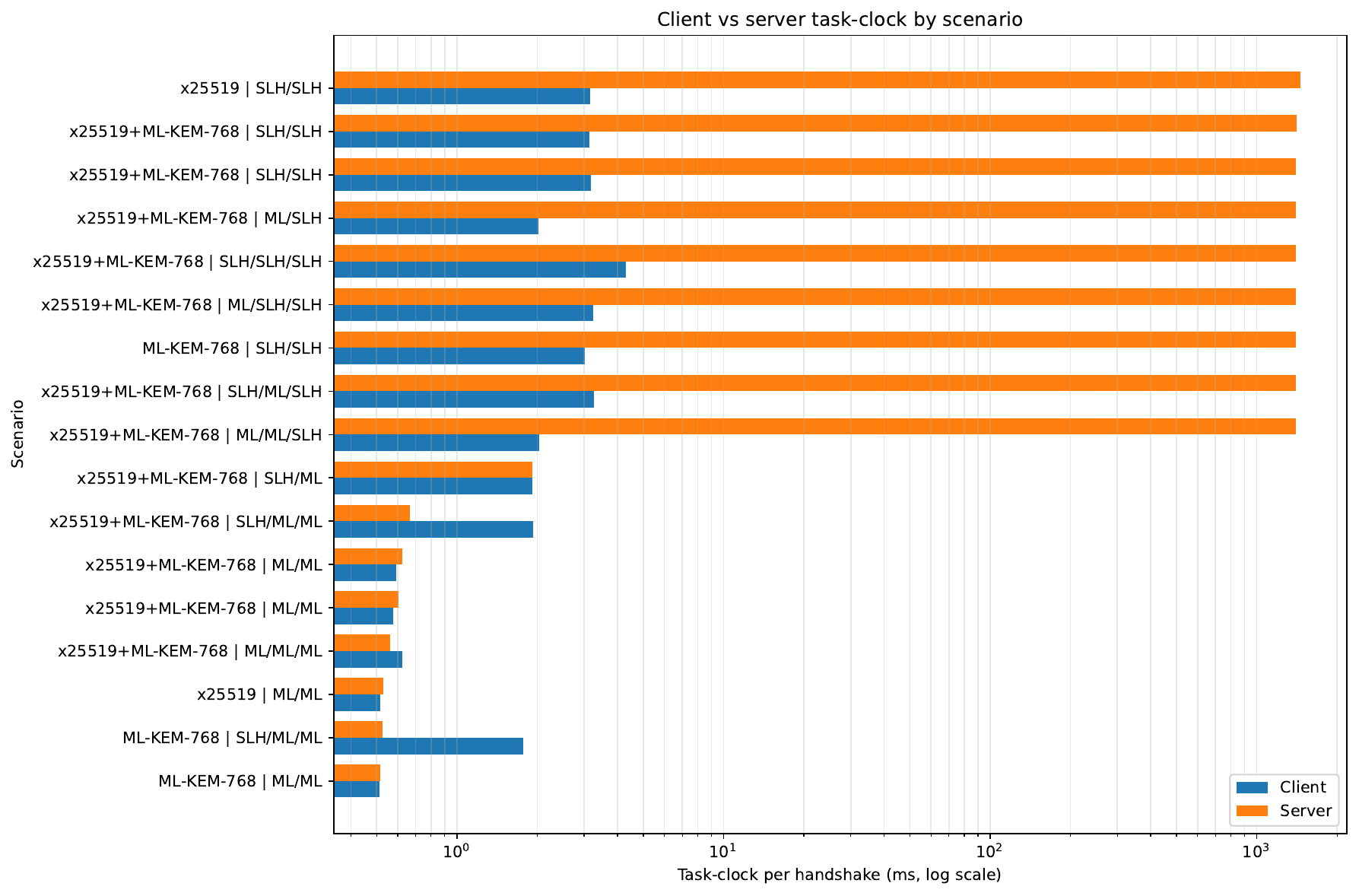}
	\caption{Client versus server task-clock by scenario.}
	\label{fig:fig09_client_vs_server_taskclock}
\end{figure}

\begin{figure}[t]
	\centering
	\includegraphics[width=0.78\linewidth]{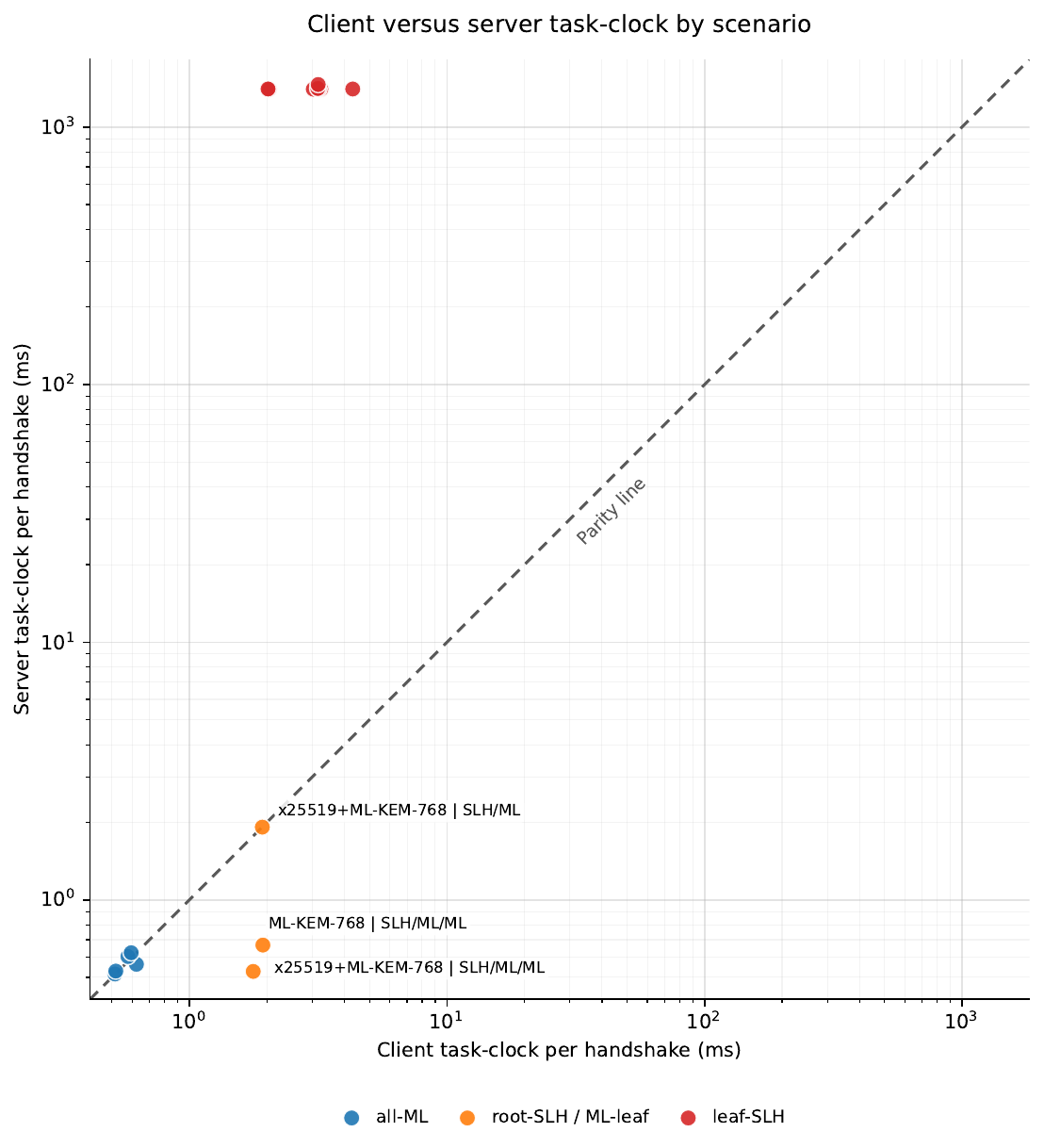}
	\caption{Client versus server task-clock by scenario on log-log axes. The diagonal marks parity between client-side and server-side active work; the clustered separation reveals distinct authentication regimes.}
	\label{fig:fig09_client_vs_server_taskclock_scatter}
\end{figure}

\begin{figure}[t]
	\centering
	\includegraphics[width=0.9\linewidth]{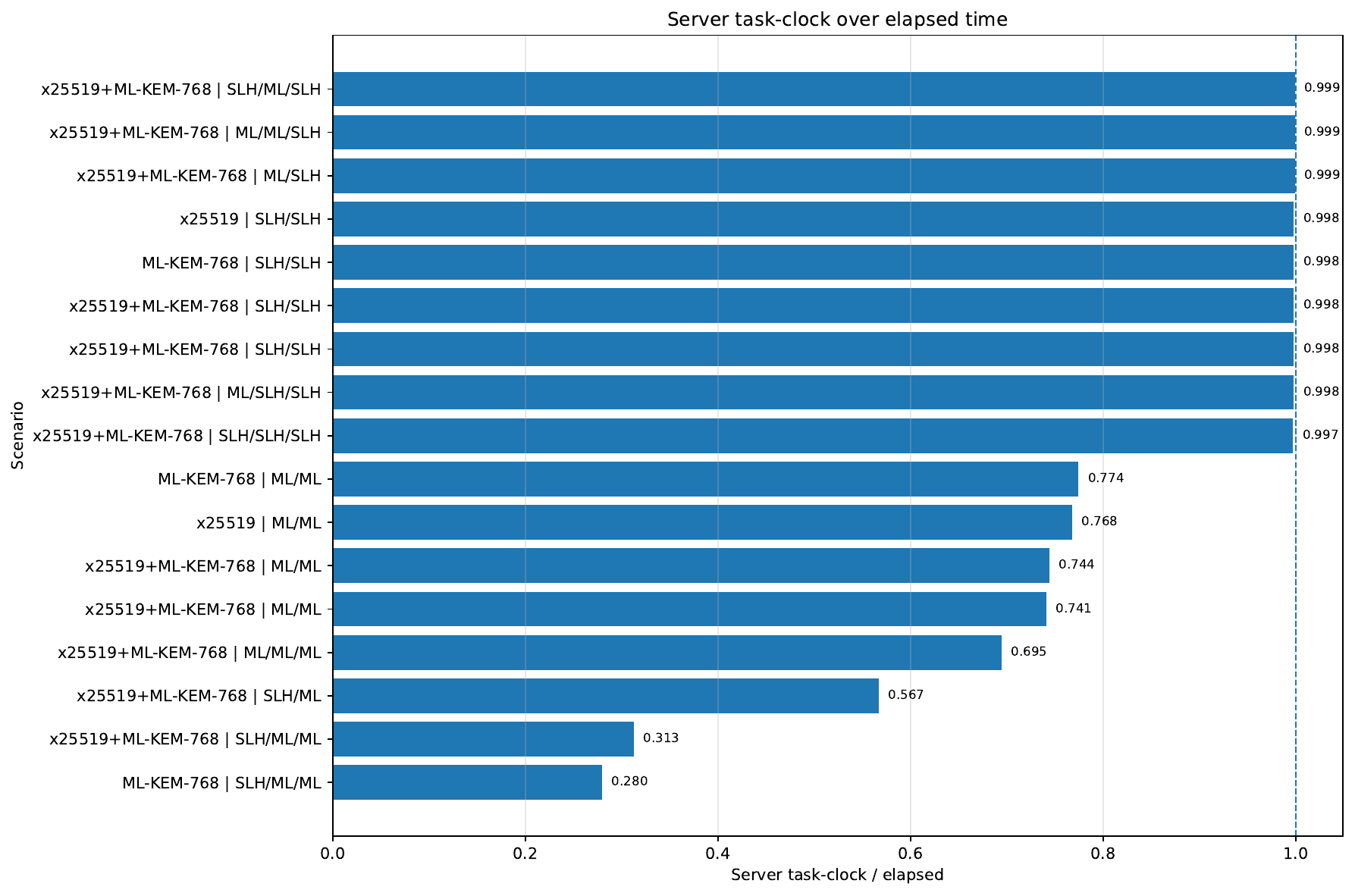}
	\caption{Server task-clock over elapsed time.}
	\label{fig:fig10_server_taskclock_over_elapsed}
\end{figure}

\begin{figure}[t]
	\centering
	\includegraphics[width=\linewidth]{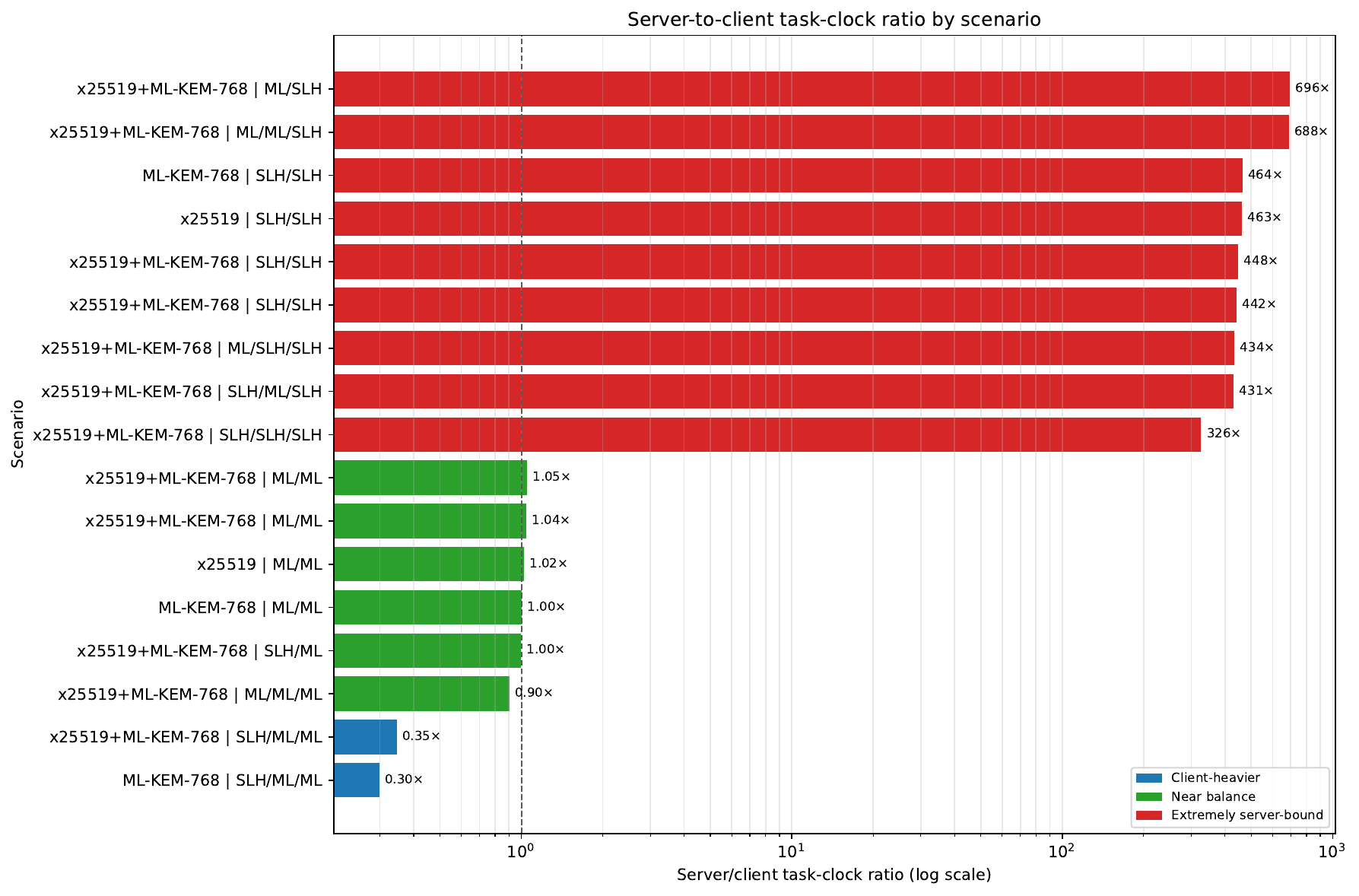}
	\caption{Server-to-client task-clock ratio by scenario.}
	\label{fig:fig17_server_client_taskclock_ratio}
\end{figure}

\begin{table}[t]
	\centering
	\caption{Performance decomposition summary.}
	\label{tab:block10_perf_decomposition_summary_body}
	\resizebox{\textwidth}{!}{%
	\begin{tabular}{lrrrr}
		\toprule
		scenario\_id & elapsed\_mean\_ms & client\_task\_clock\_per\_run\_ms & server\_task\_clock\_per\_run\_ms & client\_taskclock\_over\_elapsed \\
		\midrule
		mlkem768\_\_ml\_root\_\_ml\_leaf & 0.665200 & 0.513600 & 0.515200 & 0.772100 \\
		mlkem768\_\_slh\_root\_\_ml\_int\_\_ml\_leaf & 1.884200 & 1.767500 & 0.527300 & 0.938100 \\
		mlkem768\_\_slh\_root\_\_slh\_leaf & 1405.455900 & 3.024100 & 1402.896000 & 0.002200 \\
		x25519\_\_leaf\_mldsa65 & 0.688500 & 0.517300 & 0.528700 & 0.751400 \\
		x25519\_\_leaf\_slhdsashake192s & 1464.932700 & 3.159200 & 1462.281300 & 0.002200 \\
		x25519mlkem768\_\_leaf\_mldsa65 & 0.840600 & 0.593700 & 0.622800 & 0.706200 \\
		x25519mlkem768\_\_leaf\_slhdsashake192s & 1413.991400 & 3.148500 & 1411.344800 & 0.002200 \\
		x25519mlkem768\_\_ml\_root\_\_ml\_int\_\_ml\_leaf & 0.809100 & 0.622200 & 0.561900 & 0.769000 \\
		x25519mlkem768\_\_ml\_root\_\_ml\_int\_\_slh\_leaf & 1402.486200 & 2.036700 & 1401.169400 & 0.001500 \\
		x25519mlkem768\_\_ml\_root\_\_ml\_leaf & 0.809300 & 0.576800 & 0.602200 & 0.712700 \\
		x25519mlkem768\_\_ml\_root\_\_slh\_int\_\_slh\_leaf & 1405.802800 & 3.235800 & 1403.109500 & 0.002300 \\
		x25519mlkem768\_\_ml\_root\_\_slh\_leaf & 1406.283200 & 2.018000 & 1404.926900 & 0.001400 \\
		x25519mlkem768\_\_slh\_root\_\_ml\_int\_\_ml\_leaf & 2.133000 & 1.927300 & 0.667200 & 0.903600 \\
		x25519mlkem768\_\_slh\_root\_\_ml\_int\_\_slh\_leaf & 1403.165900 & 3.254300 & 1401.849300 & 0.002300 \\
		x25519mlkem768\_\_slh\_root\_\_ml\_leaf & 3.376200 & 1.917600 & 1.914600 & 0.568000 \\
		x25519mlkem768\_\_slh\_root\_\_slh\_int\_\_slh\_leaf & 1408.703300 & 4.303500 & 1404.903400 & 0.003100 \\
		x25519mlkem768\_\_slh\_root\_\_slh\_leaf & 1407.714100 & 3.182100 & 1405.087400 & 0.002300 \\
		\bottomrule
	\end{tabular}%
}

\vspace{0.5em}

\resizebox{\textwidth}{!}{%
	\begin{tabular}{lrrl}
		\toprule
		scenario\_id & server\_taskclock\_over\_elapsed & server\_client\_taskclock\_ratio & qualitative\_perf\_regime \\
		\midrule
		mlkem768\_\_ml\_root\_\_ml\_leaf & 0.774500 & 1.003000 & balanced \\
		mlkem768\_\_slh\_root\_\_ml\_int\_\_ml\_leaf & 0.279800 & 0.298300 & client\_skewed \\
		mlkem768\_\_slh\_root\_\_slh\_leaf & 0.998200 & 463.910400 & overwhelmingly\_server\_bound \\
		x25519\_\_leaf\_mldsa65 & 0.768000 & 1.022100 & balanced \\
		x25519\_\_leaf\_slhdsashake192s & 0.998200 & 462.869300 & overwhelmingly\_server\_bound \\
		x25519mlkem768\_\_leaf\_mldsa65 & 0.741000 & 1.049200 & balanced \\
		x25519mlkem768\_\_leaf\_slhdsashake192s & 0.998100 & 448.264200 & overwhelmingly\_server\_bound \\
		x25519mlkem768\_\_ml\_root\_\_ml\_int\_\_ml\_leaf & 0.694500 & 0.903100 & balanced \\
		x25519mlkem768\_\_ml\_root\_\_ml\_int\_\_slh\_leaf & 0.999100 & 687.949400 & overwhelmingly\_server\_bound \\
		x25519mlkem768\_\_ml\_root\_\_ml\_leaf & 0.744100 & 1.044100 & balanced \\
		x25519mlkem768\_\_ml\_root\_\_slh\_int\_\_slh\_leaf & 0.998100 & 433.616100 & overwhelmingly\_server\_bound \\
		x25519mlkem768\_\_ml\_root\_\_slh\_leaf & 0.999000 & 696.186200 & overwhelmingly\_server\_bound \\
		x25519mlkem768\_\_slh\_root\_\_ml\_int\_\_ml\_leaf & 0.312800 & 0.346200 & client\_skewed \\
		x25519mlkem768\_\_slh\_root\_\_ml\_int\_\_slh\_leaf & 0.999100 & 430.763900 & overwhelmingly\_server\_bound \\
		x25519mlkem768\_\_slh\_root\_\_ml\_leaf & 0.567100 & 0.998500 & balanced \\
		x25519mlkem768\_\_slh\_root\_\_slh\_int\_\_slh\_leaf & 0.997300 & 326.456000 & overwhelmingly\_server\_bound \\
		x25519mlkem768\_\_slh\_root\_\_slh\_leaf & 0.998100 & 441.559800 & overwhelmingly\_server\_bound \\
		\bottomrule
	\end{tabular}%
}

\end{table}

\subsection{Limits of client-only observations}

Earlier stages of the TLS benchmarking work suggested that client-side activity could not account for the second-scale latency observed in the heaviest scenarios. Client-only observations were nevertheless incomplete: they could indicate that the client did not consume enough active CPU time to justify the total elapsed duration, but they could not identify where the remaining time was spent.

A slow handshake may result from heavier client-side validation, heavier server-side authentication work, roughly symmetric stress on both peers, or waiting behavior dominated by one peer while the other remains active. Without server-side measurements, those possibilities remain partially entangled. The inclusion of server-side performance counters in the final dataset closes that gap and allows the heavy regime to be interpreted in workload terms rather than only in latency terms.

The decomposition distinguishes among three possible structures for the leaf-SLH regime: balanced work across client and server, a shift toward client-side validation, or overwhelming concentration of active work on the server side.

\subsection{Balanced regime in all-ML scenarios}

The all-ML class remains the reference regime in both latency and workload structure. Across those scenarios, mean client task-clock is approximately 0.5647~ms, while mean server task-clock is approximately 0.5662~ms. Normalized by elapsed time, the client contributes about 0.7423 of elapsed time in active work and the server about 0.7444. The mean server/client task-clock ratio is approximately 1.0043, and the mean server/client instruction ratio is approximately 1.0841.

These values describe a regime that is low-cost and structurally balanced. Both ends of the handshake perform comparable amounts of active work, and neither dominates the other. Figure~\ref{fig:fig09_client_vs_server_taskclock} shows this directly: in the low-cost all-ML region, client and server bars remain of similar magnitude. Figure~\ref{fig:fig09_client_vs_server_taskclock_scatter} shows the same structure geometrically, with all-ML scenarios clustering close to the parity line. Figure~\ref{fig:fig17_server_client_taskclock_ratio} reinforces the point through ratios close to one.

This balanced structure provides the control case against which the remaining regimes can be interpreted. Without it, the heavy scenarios might be read simply as more expensive instances of the same general phenomenon. The decomposition shows that they differ qualitatively in where active work resides.

\subsection{Validation-skewed regime in upper-layer SLH scenarios}

The next distinct regime is formed by scenarios in which SLH-DSA appears in upper trust layers while the interactive leaf remains ML-DSA. These cases incur a visible performance penalty while remaining operationally plausible, and their decomposition shows that the penalty is not distributed symmetrically.

For the \texttt{root-SLH / leaf-ML} class, mean client task-clock rises to approximately 1.8708~ms, whereas mean server task-clock rises only to about 1.0364~ms. When normalized by elapsed time, the client contributes approximately 0.8032 of elapsed time in active work, while the server contributes only about 0.3866. The mean server/client task-clock ratio falls to approximately 0.5477, and the mean instruction ratio to about 0.4360.

This is a validation-skewed regime rather than a server-collapse pattern. The server does more work than in the all-ML baseline, but the relative burden shifts more strongly toward the client. The most natural interpretation is that placing SLH-DSA above the leaf increases the validation burden visible on the client side without pushing the interactive server-authentication step into the catastrophic regime.

The contrast with the full strategy matrix makes this distinction clear. The \\
\texttt{x25519mlkem768\_\_slh\_root\_\_ml\_int\_\_ml\_leaf} strategy is materially slower than the fully-ML baseline, but it does not resemble the leaf-SLH cases in workload structure. Geometrically, these scenarios move away from the parity cluster without entering the vertically separated server-dominated cloud shown in Figure~\ref{fig:fig09_client_vs_server_taskclock_scatter}. That difference indicates that upper-layer SLH and leaf-SLH are not weaker and stronger versions of the same phenomenon.

\subsection{Overwhelmingly server-bound regime in leaf-SLH scenarios}

The decomposition changes sharply once SLH-DSA reaches the leaf. In that class, mean client task-clock is only about 3.0402~ms, while mean server task-clock rises to approximately 1410.8409~ms. The normalized ratios are even more informative: the client contributes only about 0.0022 of elapsed time in active work, whereas the server contributes about 0.9984. The mean server/client task-clock ratio is approximately 487.95, and the mean server/client instruction ratio is approximately 648.91.

At that point the handshake becomes overwhelmingly server-bound. Figure~\ref{fig:fig08_server_taskclock_by_scenario} shows that server-side task-clock grows sharply only in the leaf-SLH cases. Figure~\ref{fig:fig09_client_vs_server_taskclock_scatter} shows the corresponding geometric separation: leaf-SLH scenarios form a distinct cloud far above the client/server parity line rather than a gradual extension of the low-cost region. Figure~\ref{fig:fig10_server_taskclock_over_elapsed} makes the same point directly. In those scenarios, server task-clock almost coincides with end-to-end elapsed time. In practical terms, the total handshake duration is almost entirely accounted for by server-side active compute, while the client is left largely in a waiting role.

Figure~\ref{fig:fig17_server_client_taskclock_ratio} summarizes the same asymmetry. Ratios in leaf-SLH scenarios jump from near-unity or sub-unity values into the hundreds. Some of the most extreme cases show server/client task-clock ratios above 680 and instruction ratios above 1000. The handshake is dominated by server-side work to an extent that marks a clear regime change.

Among the most server-dominated scenarios are \texttt{x25519mlkem768\_\_slh\_root\_\_ml\_int\_\_slh\_leaf}, \texttt{x25519mlkem768\_\_ml\_root\_\_ml\_int\_\_slh\_leaf}, and \texttt{x25519mlkem768\_\_ml\_root\_\_slh\_leaf}, all of which exhibit \texttt{server\_taskclock\_over\_elapsed} ratios around 0.999 together with very large server/client asymmetries. The decomposition therefore sharpens the central claim of the paper into a causal statement. The degradation observed in leaf-SLH scenarios is neither primarily a transport phenomenon nor primarily a client-validation phenomenon. It is overwhelmingly a server-side cryptographic compute phenomenon concentrated in the live authentication path.

\subsection{Supporting microarchitectural signals}

The dataset also includes lower-level counters such as IPC, cache-miss rate, and branch-miss rate. The present study is not a function-tracing or microarchitectural reverse-engineering exercise, and these counters are used only as supporting evidence for regime differentiation rather than as proof of a unique internal mechanism.

With that caution in place, the counters still suggest that the heavy leaf-SLH regime is not simply more of the same workload. In the low-cost ML scenarios, the server tends to operate in a comparatively modest regime, with IPC values around 2.1--2.35 and cache-miss rates in the low single-digit range. By contrast, the heavy leaf-SLH scenarios exhibit a stable and sharply separated profile, with server IPC around 4.54--4.55, cache-miss rates roughly in the 33--40\% range, and branch-miss rates around 0.48--0.51\%.

The paper does not identify a single internal OpenSSL or provider function as the unique source of cost, nor does it reconstruct the complete internal execution path. The defensible point is that the leaf-SLH cases exhibit a stable server-side execution profile that is sharply distinct from the low-cost ML regime.

Taken together, the decomposition results complete the causal arc of the paper. The previous section showed that leaf-SLH defines a distinct authentication regime that transport alone does not explain. The present section shows why: once SLH-DSA reaches the server leaf, the handshake becomes almost entirely a server-side compute event. That finding provides the bridge from hierarchy design to deployment consequence and prepares the next step of the argument, namely the translation of authentication cost into capacity loss and operational viability.

\section{Operational Implications for PQ TLS Deployment}

The previous sections showed that the dominant heavy regime is overwhelmingly server-side and concentrated in leaf-SLH scenarios. The operational question is how second-scale server-authentication cost should be translated into capacity loss, infrastructure scaling, and deployment plausibility for a service operator.

\begin{figure}[t]
	\centering
	\includegraphics[width=\linewidth]{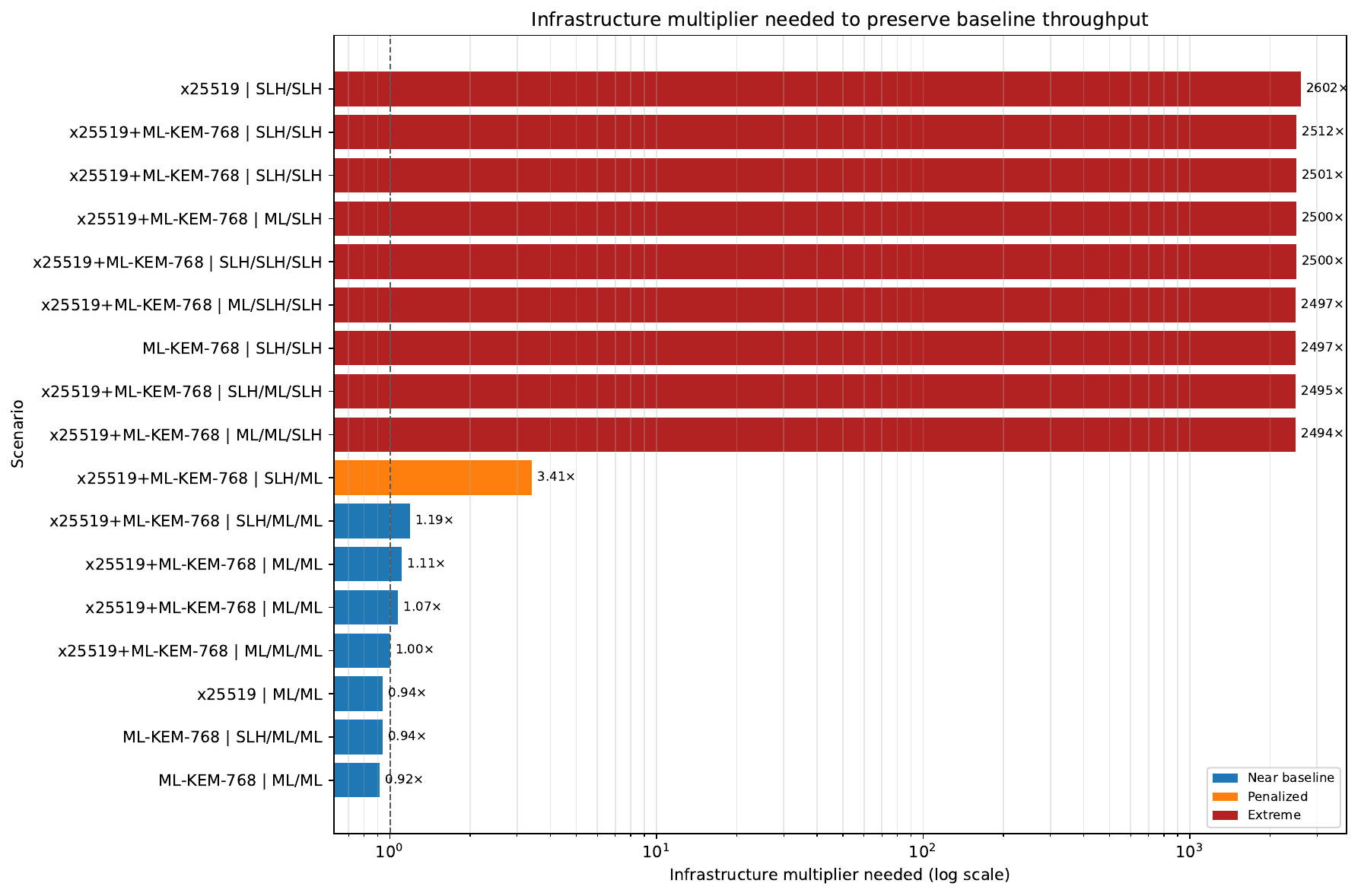}
	\caption{Infrastructure multiplier needed to preserve baseline throughput.}
	\label{fig:fig14_infrastructure_multiplier_needed}
\end{figure}

\begin{figure}[t]
	\centering
	\includegraphics[width=\linewidth]{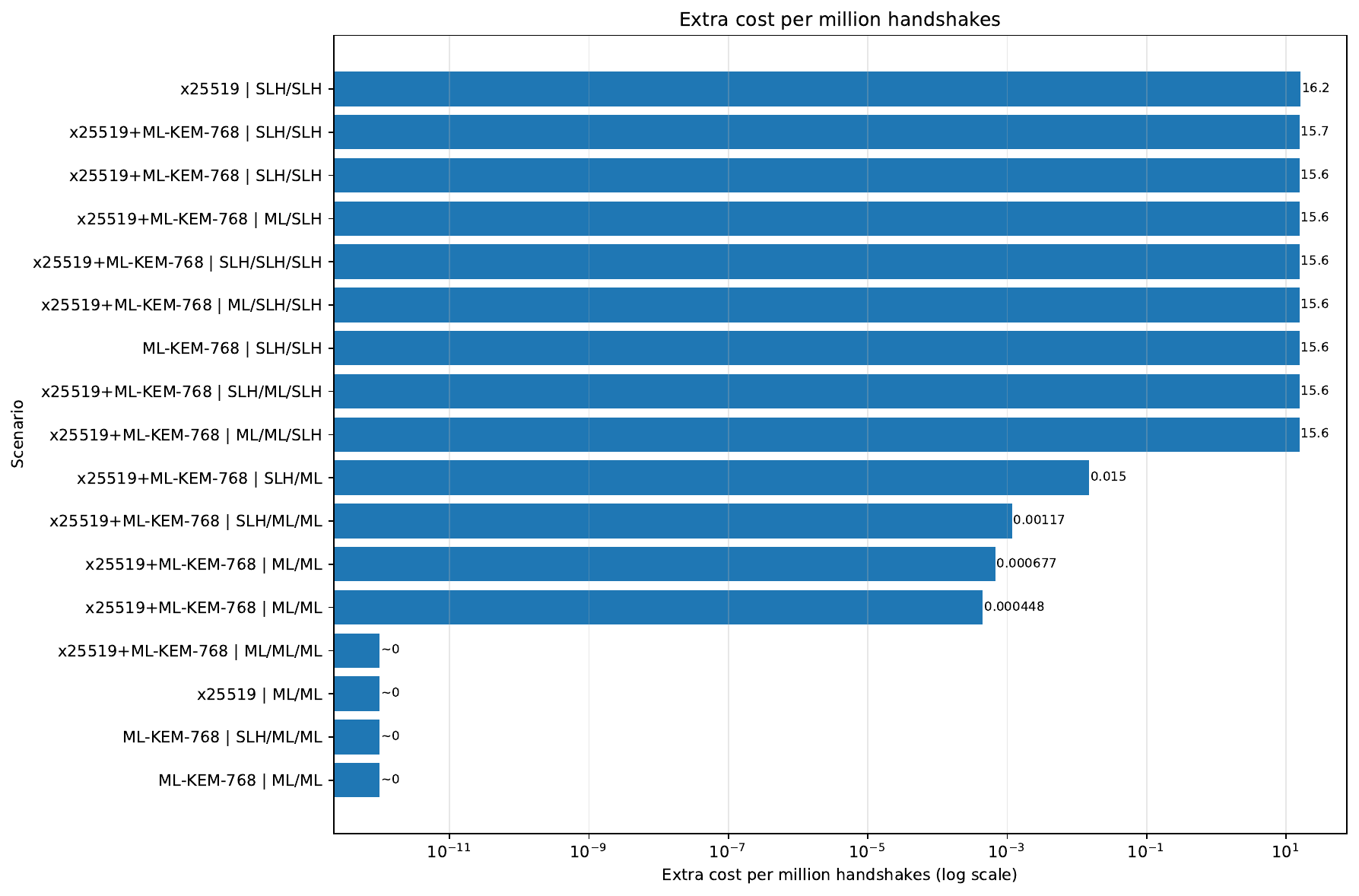}
	\caption{Extra cost per million handshakes.}
	\label{fig:fig15_extra_cost_per_million_handshakes}
\end{figure}

\begin{figure}[t]
	\centering
	\includegraphics[width=0.95\linewidth]{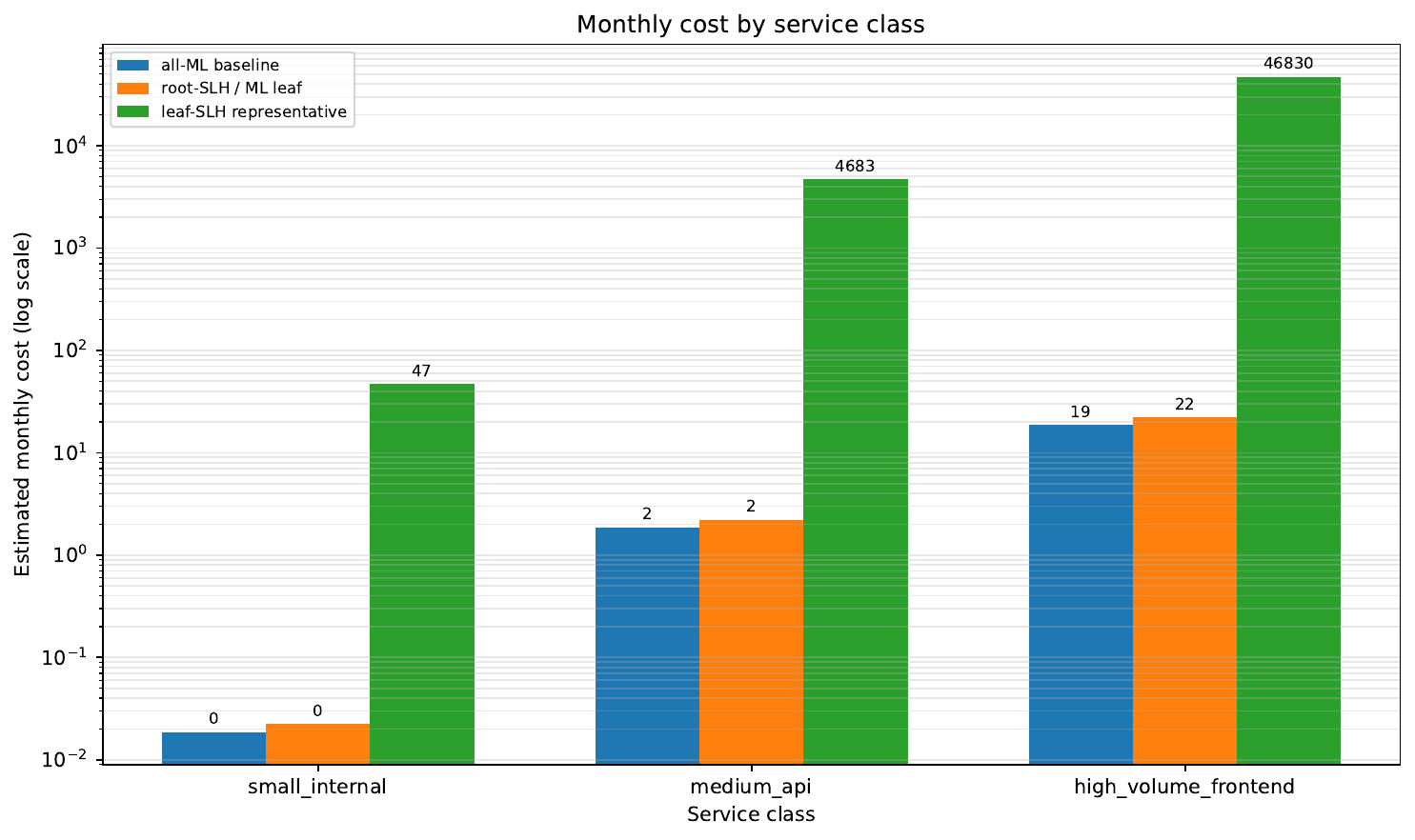}
	\caption{Monthly cost by service class.}
	\label{fig:fig16_monthly_cost_by_service_class}
\end{figure}

\begin{table}[t]
	\centering
	\caption{Infrastructure capacity summary.}
	\label{tab:block11_1_capacity_model_body}
	\resizebox{\textwidth}{!}{%
	\begin{tabular}{llrrrr}
		\toprule
		scenario\_id & kex\_mode & depth & handshakes\_per\_core\_second & handshakes\_per\_vcpu\_hour & capacity\_retained\_vs\_baseline \\
		\midrule
		mlkem768\_\_ml\_root\_\_ml\_leaf & pure\_pqc & 2 & 1,940.97 & 6,987,482.70 & 1.0906 \\
		mlkem768\_\_slh\_root\_\_ml\_int\_\_ml\_leaf & pure\_pqc & 3 & 1,896.59 & 6,827,725.12 & 1.0657 \\
		x25519\_\_leaf\_mldsa65 & classical & 2 & 1,891.31 & 6,808,716.67 & 1.0627 \\
		x25519mlkem768\_\_ml\_root\_\_ml\_int\_\_ml\_leaf & hybrid & 3 & 1,779.68 & 6,406,856.76 & 1.0000 \\
		x25519mlkem768\_\_ml\_root\_\_ml\_leaf & hybrid & 2 & 1,660.56 & 5,978,030.74 & 0.9331 \\
		x25519mlkem768\_\_leaf\_mldsa65 & hybrid & 2 & 1,605.53 & 5,779,919.92 & 0.9021 \\
		x25519mlkem768\_\_slh\_root\_\_ml\_int\_\_ml\_leaf & hybrid & 3 & 1,498.86 & 5,395,885.64 & 0.8422 \\
		x25519mlkem768\_\_slh\_root\_\_ml\_leaf & hybrid & 2 & 522.29 & 1,880,240.19 & 0.2935 \\
		x25519mlkem768\_\_ml\_root\_\_ml\_int\_\_slh\_leaf & hybrid & 3 & 0.71 & 2,569.28 & 0.0004 \\
		x25519mlkem768\_\_slh\_root\_\_ml\_int\_\_slh\_leaf & hybrid & 3 & 0.71 & 2,568.04 & 0.0004 \\
		mlkem768\_\_slh\_root\_\_slh\_leaf & pure\_pqc & 2 & 0.71 & 2,566.12 & 0.0004 \\
		x25519mlkem768\_\_ml\_root\_\_slh\_int\_\_slh\_leaf & hybrid & 3 & 0.71 & 2,565.73 & 0.0004 \\
		x25519mlkem768\_\_slh\_root\_\_slh\_int\_\_slh\_leaf & hybrid & 3 & 0.71 & 2,562.45 & 0.0004 \\
		x25519mlkem768\_\_ml\_root\_\_slh\_leaf & hybrid & 2 & 0.71 & 2,562.41 & 0.0004 \\
		x25519mlkem768\_\_slh\_root\_\_slh\_leaf & hybrid & 2 & 0.71 & 2,562.12 & 0.0004 \\
		x25519mlkem768\_\_leaf\_slhdsashake192s & hybrid & 2 & 0.71 & 2,550.76 & 0.0004 \\
		x25519\_\_leaf\_slhdsashake192s & classical & 2 & 0.68 & 2,461.91 & 0.0004 \\
		\bottomrule
	\end{tabular}%
}

\vspace{0.5em}

\resizebox{\textwidth}{!}{%
	\begin{tabular}{llrl}
		\toprule
		scenario\_id & kex\_mode & infrastructure\_multiplier\_needed & conceptual\_perf\_group \\
		\midrule
		mlkem768\_\_ml\_root\_\_ml\_leaf & pure\_pqc & 0.9169 & all\_ml \\
		mlkem768\_\_slh\_root\_\_ml\_int\_\_ml\_leaf & pure\_pqc & 0.9384 & root\_slh\_leaf\_ml \\
		x25519\_\_leaf\_mldsa65 & classical & 0.9410 & all\_ml \\
		x25519mlkem768\_\_ml\_root\_\_ml\_int\_\_ml\_leaf & hybrid & 1.0000 & all\_ml \\
		x25519mlkem768\_\_ml\_root\_\_ml\_leaf & hybrid & 1.0717 & all\_ml \\
		x25519mlkem768\_\_leaf\_mldsa65 & hybrid & 1.1085 & all\_ml \\
		x25519mlkem768\_\_slh\_root\_\_ml\_int\_\_ml\_leaf & hybrid & 1.1874 & root\_slh\_leaf\_ml \\
		x25519mlkem768\_\_slh\_root\_\_ml\_leaf & hybrid & 3.4075 & root\_slh\_leaf\_ml \\
		x25519mlkem768\_\_ml\_root\_\_ml\_int\_\_slh\_leaf & hybrid & 2493.6366 & leaf\_slh \\
		x25519mlkem768\_\_slh\_root\_\_ml\_int\_\_slh\_leaf & hybrid & 2494.8466 & leaf\_slh \\
		mlkem768\_\_slh\_root\_\_slh\_leaf & pure\_pqc & 2496.7094 & leaf\_slh \\
		x25519mlkem768\_\_ml\_root\_\_slh\_int\_\_slh\_leaf & hybrid & 2497.0893 & leaf\_slh \\
		x25519mlkem768\_\_slh\_root\_\_slh\_int\_\_slh\_leaf & hybrid & 2500.2820 & leaf\_slh \\
		x25519mlkem768\_\_ml\_root\_\_slh\_leaf & hybrid & 2500.3237 & leaf\_slh \\
		x25519mlkem768\_\_slh\_root\_\_slh\_leaf & hybrid & 2500.6093 & leaf\_slh \\
		x25519mlkem768\_\_leaf\_slhdsashake192s & hybrid & 2511.7455 & leaf\_slh \\
		x25519\_\_leaf\_slhdsashake192s & classical & 2602.3964 & leaf\_slh \\
		\bottomrule
	\end{tabular}%
}

\end{table}

\begin{table}[t]
	\centering
	\small
	\setlength{\tabcolsep}{4pt}
	\renewcommand{\arraystretch}{1.08}
	\caption{Economic summary by scenario/class.}
	\label{tab:block11_2_economic_model_body}
	\resizebox{\textwidth}{!}{%
	\begin{tabular}{L{4.0cm} L{2.7cm} r r r r r r}
		\toprule
		\makecell[l]{scenario\_id} &
		\makecell[l]{conceptual\\economic class} &
		\makecell[r]{server CPU seconds\\per handshake} &
		\makecell[r]{handshakes per\\server CPU second} &
		\makecell[r]{handshakes per\\server CPU hour} &
		\makecell[r]{capacity retained\\vs baseline} &
		\makecell[r]{capacity loss\\vs baseline (\%)} &
		\makecell[r]{infrastructure\\multiplier needed} \\
		\midrule
		\seqsplit{x25519mlkem768\_\_ml\_root\_\_ml\_int\_\_ml\_leaf} & all\_ml & 0.0006 & 1779.6824 & 6406856.7605 & 1.0000 & 0.0000 & 1.0000 \\
		\seqsplit{x25519mlkem768\_\_slh\_root\_\_ml\_int\_\_ml\_leaf} & root\_slh\_leaf\_ml & 0.0007 & 1498.8571 & 5395885.6372 & 0.8422 & 15.7795 & 1.1874 \\
		\seqsplit{x25519mlkem768\_\_slh\_root\_\_slh\_int\_\_slh\_leaf} & leaf\_slh & 1.4049 & 0.7118 & 2562.4537 & 0.0004 & 99.9600 & 2500.2820 \\
		\seqsplit{x25519mlkem768\_\_ml\_root\_\_slh\_leaf} & leaf\_slh & 1.4049 & 0.7118 & 2562.4109 & 0.0004 & 99.9600 & 2500.3237 \\
		\seqsplit{x25519\_\_leaf\_slhdsashake192s} & leaf\_slh & 1.4623 & 0.6839 & 2461.9066 & 0.0004 & 99.9616 & 2602.3964 \\
		\bottomrule
	\end{tabular}%
}

\vspace{0.5em}

\resizebox{\textwidth}{!}{%
	\begin{tabular}{L{4.0cm} L{2.7cm} r r r r}
		\toprule
		\makecell[l]{scenario\_id} &
		\makecell[l]{conceptual\\economic class} &
		\makecell[r]{CPU hours per\\million handshakes} &
		\makecell[r]{cost per million\\default} &
		\makecell[r]{extra cost per\\million default} &
		\makecell[r]{cost multiplier\\vs baseline} \\
		\midrule
		\seqsplit{x25519mlkem768\_\_ml\_root\_\_ml\_int\_\_ml\_leaf} & all\_ml & 0.1561 & 0.0062 & 0.0000 & 1.0000 \\
		\seqsplit{x25519mlkem768\_\_slh\_root\_\_ml\_int\_\_ml\_leaf} & root\_slh\_leaf\_ml & 0.1853 & 0.0074 & 0.0012 & 1.1874 \\
		\seqsplit{x25519mlkem768\_\_slh\_root\_\_slh\_int\_\_slh\_leaf} & leaf\_slh & 390.2510 & 15.6100 & 15.6038 & 2500.2820 \\
		\seqsplit{x25519mlkem768\_\_ml\_root\_\_slh\_leaf} & leaf\_slh & 390.2575 & 15.6103 & 15.6041 & 2500.3237 \\
		\seqsplit{x25519\_\_leaf\_slhdsashake192s} & leaf\_slh & 406.1893 & 16.2476 & 16.2413 & 2602.3964 \\
		\bottomrule
	\end{tabular}%
}
\end{table}

\begin{table}[t]
	\centering
	\small
	\setlength{\tabcolsep}{4pt}
	\renewcommand{\arraystretch}{1.08}
	\caption{Service-class economic translation.}
	\label{tab:block11_2_economic_model_service_class_summary}
	\resizebox{\textwidth}{!}{%
	\begin{tabular}{L{3.0cm} L{2.8cm} r r r r r r r}
		\toprule
		\makecell[l]{service\\class} &
		\makecell[l]{conceptual\\economic class} &
		\makecell[c]{scenarios} &
		\makecell[r]{mean daily\\cost} &
		\makecell[r]{median daily\\cost} &
		\makecell[r]{mean extra daily cost\\vs baseline} &
		\makecell[r]{median extra daily cost\\vs baseline} &
		\makecell[r]{mean monthly\\cost (30d)} &
		\makecell[r]{median monthly\\cost (30d)} \\
		\midrule
		high\_volume\_frontend & all\_ml & 5 & 0.6291 & 0.6243 & 0.0048 & 0.0000 & 18.8726 & 18.7299 \\
		high\_volume\_frontend & leaf\_slh & 9 & 1567.6010 & 1561.0038 & 1566.9767 & 1560.3795 & 47028.0298 & 46830.1144 \\
		high\_volume\_frontend & root\_slh\_leaf\_ml & 3 & 1.1515 & 0.7413 & 0.5272 & 0.1170 & 34.5454 & 22.2392 \\
		medium\_api & all\_ml & 5 & 0.0629 & 0.0624 & 0.0005 & 0.0000 & 1.8873 & 1.8730 \\
		medium\_api & leaf\_slh & 9 & 156.7601 & 156.1004 & 156.6977 & 156.0379 & 4702.8030 & 4683.0114 \\
		medium\_api & root\_slh\_leaf\_ml & 3 & 0.1152 & 0.0741 & 0.0527 & 0.0117 & 3.4545 & 2.2239 \\
		small\_internal & all\_ml & 5 & 0.0006 & 0.0006 & 0.0000 & 0.0000 & 0.0189 & 0.0187 \\
		small\_internal & leaf\_slh & 9 & 1.5676 & 1.5610 & 1.5670 & 1.5604 & 47.0280 & 46.8301 \\
		small\_internal & root\_slh\_leaf\_ml & 3 & 0.0012 & 0.0007 & 0.0005 & 0.0001 & 0.0345 & 0.0222 \\
		\bottomrule
	\end{tabular}%
}

\vspace{0.5em}

\resizebox{\textwidth}{!}{%
	\begin{tabular}{L{3.0cm} L{2.8cm} r r r r r}
		\toprule
		\makecell[l]{service\\class} &
		\makecell[l]{conceptual\\economic class} &
		\makecell[c]{scenarios} &
		\makecell[r]{mean extra monthly cost\\(30d) vs baseline} &
		\makecell[r]{median extra monthly cost\\(30d) vs baseline} &
		\makecell[r]{mean annual\\cost (365d)} &
		\makecell[r]{mean extra annual cost\\(365d) vs baseline} \\
		\midrule
		high\_volume\_frontend & all\_ml & 5 & 0.1427 & 0.0000 & 229.6166 & 1.7358 \\
		high\_volume\_frontend & leaf\_slh & 9 & 47009.2998 & 46811.3845 & 572174.3620 & 571946.4811 \\
		high\_volume\_frontend & root\_slh\_leaf\_ml & 3 & 15.8155 & 3.5092 & 420.3024 & 192.4215 \\
		medium\_api & all\_ml & 5 & 0.0143 & 0.0000 & 22.9617 & 0.1736 \\
		medium\_api & leaf\_slh & 9 & 4700.9300 & 4681.1385 & 57217.4362 & 57194.6481 \\
		medium\_api & root\_slh\_leaf\_ml & 3 & 1.5815 & 0.3509 & 42.0302 & 19.2422 \\
		small\_internal & all\_ml & 5 & 0.0001 & 0.0000 & 0.2296 & 0.0017 \\
		small\_internal & leaf\_slh & 9 & 47.0093 & 46.8114 & 572.1744 & 571.9465 \\
		small\_internal & root\_slh\_leaf\_ml & 3 & 0.0158 & 0.0035 & 0.4203 & 0.1924 \\
		\bottomrule
	\end{tabular}%
}
\end{table}

\subsection{From server-side handshake cost to service capacity}

For an interactive TLS service, handshake cost is both a latency issue and a capacity issue. Every additional unit of server-side work per handshake reduces the number of sessions a node can sustain over time, lowers effective throughput per core, and increases the infrastructure required to maintain a fixed service rate. In high-volume settings, these effects accumulate quickly. A per-handshake cryptographic penalty therefore becomes a constraint on concurrency, scaling, and operational headroom.

The interpretation must therefore go beyond raw latency. Once the client/server decomposition showed that the heavy regime is overwhelmingly server-side, the relevant unit of interpretation became server CPU time per handshake. The translation to service capacity then follows directly: if one authentication strategy consumes orders of magnitude more server compute per handshake than another, throughput per node falls proportionally, and sustaining the same rate of authenticated connections requires proportionally more capacity.

Under that view, the heavy leaf-SLH regime is an authentication design that consumes server compute in a way that directly constrains deployable scale. For an operator, the relevant question is whether the handshake remains viable at the service layer.

\subsection{Capacity loss under leaf-SLH}

The most direct operational translation of the server-side measurements is handshake capacity. Table~\ref{tab:block11_1_capacity_model_body} reports derived capacity metrics such as handshakes per core-second, handshakes per vCPU-hour, retained capacity relative to baseline, and the infrastructure multiplier required to preserve baseline throughput.

The hybrid depth-3 fully-ML baseline,
\texttt{x25519mlkem768\_\_ml\_root\_\_ml\_int\_\_ml\_leaf},
supports approximately 1779.68 handshakes per core-second, or about 6,406,856.76 handshakes per vCPU-hour. This is the reference point against which the remaining strategies are interpreted.

The all-ML family remains essentially aligned with that baseline. Its mean retained capacity is approximately 0.9977$\times$ baseline, and its mean infrastructure multiplier is approximately 1.0076$\times$. In practical terms, the all-ML region is capacity-stable.

The \texttt{root-SLH / leaf-ML} family introduces a real but bounded degradation. Its mean retained capacity is approximately 0.7338$\times$ baseline, implying a mean infrastructure multiplier of approximately 1.8444$\times$, with a much more moderate median multiplier of approximately 1.1874$\times$. This is not negligible, but it remains within a range that could be tolerated in some deployment contexts, especially where a heavier upper trust layer is judged worthwhile.

The \texttt{leaf-SLH} class behaves very differently. Its mean retained capacity falls to approximately 0.0004$\times$ baseline, and its mean infrastructure multiplier reaches approximately 2510.85$\times$, with a median around 2500.28$\times$. Figure~\ref{fig:fig14_infrastructure_multiplier_needed} shows the scale of this discontinuity. The leaf-SLH scenarios stand several orders of magnitude beyond baseline.

At that point, the architectural distinction becomes operationally unavoidable. A configuration that requires on the order of $2500\times$ more infrastructure to preserve baseline throughput cannot reasonably be described only as less efficient. It belongs to a different deployment class altogether. The language of bounded penalty no longer applies; the observed effect is a collapse of handshake throughput as a practical front-end primitive.

\subsection{Cost-per-million-handshakes interpretation}

Capacity loss is already an infrastructure result, but cost per million handshakes provides a second operational translation. This metric is portable across traffic scales, easy to compare, and tied directly to per-handshake server burden.

Table~\ref{tab:block11_2_economic_model_body} and Figure~\ref{fig:fig15_extra_cost_per_million_handshakes} summarize this view. Under the illustrative price model used in the analysis, the fully-ML baseline costs approximately 0.006243 per million handshakes. The \texttt{root-SLH / ml\_int / ml\_leaf} strategy rises only modestly, to approximately 0.007413 per million, for an extra cost of about 0.001170 and a cost multiplier of about 1.187$\times$.

The leaf-SLH cases expand economically in direct parallel with their compute behavior. Representative heavy scenarios such as
\texttt{x25519mlkem768\_\_slh\_root\_\_slh\_int\_\_slh\_leaf}
and
\texttt{x25519mlkem768\_\_ml\_root\_\_slh\_leaf}
land around 15.61 cost units per million handshakes, yielding extra costs above 15.60 relative to baseline and cost multipliers near $2500\times$. The most extreme observed case,
\texttt{x25519\_\_leaf\_slhdsashake192s},
reaches approximately 16.247570 per million handshakes, or about 2602.396$\times$ the baseline.

This metric compresses the operational meaning of measured server-side cost into a quantity that remains intelligible across contexts. In that representation, upper-layer SLH appears as a bounded premium. Leaf-SLH appears as a structural departure from the baseline rather than as an expensive ordinary variant.

\subsection{Service-class translation}

Cost per million handshakes is portable, but operators often reason in deployment classes rather than in normalized per-million metrics. For that reason, the analysis also expresses the economic model through three illustrative service classes:
\texttt{small\_internal},
\texttt{medium\_api},
and
\texttt{high\_volume\_frontend}.

Table~\ref{tab:block11_2_economic_model_service_class_summary} and Figure~\ref{fig:fig16_monthly_cost_by_service_class} present that translation. Under the illustrative assumptions of the model, the median extra monthly cost of the \texttt{root-SLH / leaf-ML} class remains small: around 0.0035 for a small internal service, around 0.3509 for a medium API, and around 3.5092 for a high-volume front-end. This is consistent with the bounded-penalty interpretation developed in the earlier sections.

The leaf-SLH class behaves very differently. Its median extra monthly cost rises from approximately 46.8114 in the small internal service class to about 4681.1385 in the medium API class and about 46811.3845 in the high-volume front-end class. The scaling pattern itself is unsurprising, since the model is linear in traffic. In practical terms, a decision that may look merely expensive in a low-volume setting becomes economically severe in an edge-facing or high-throughput interactive deployment.

\subsection{Operational plausibility of hierarchy strategies}

The final step is to restate the results in explicitly operational terms. Earlier in the paper, the strategy space was described through four classes:
\emph{Reasonable},
\emph{Penalized but plausible},
\emph{Operationally problematic},
and
\emph{Unsuitable for interactive TLS front-end}.

The all-ML strategies remain firmly in the \emph{Reasonable} class. They are low-latency, structurally balanced, capacity-stable, and economically aligned with the baseline. The \texttt{root-SLH /} \\ \texttt{intermediate-ML / leaf-ML} strategy belongs in \emph{Penalized but plausible}. It introduces a real cost in latency, capacity, and deployment overhead, but remains within a bounded interactive regime.

The leaf-SLH strategies do not belong to the same continuum. Their latency plateau, server-bound compute profile, capacity collapse, infrastructure multipliers, and cost expansion all point in the same direction. In low-volume settings they may still be tolerated for reasons outside performance, but from the perspective of interactive service engineering they belong in the \emph{Unsuitable} class. The heavy regime is operationally disqualifying for front-end TLS use.

\section{Threats to Validity and Limitations}

The conclusions of this paper are bounded by the experimental setting studied here. Their scope is a controlled local TLS laboratory, a specific implementation stack, and a defined set of post-quantum certificate-hierarchy scenarios. Several limitations therefore constrain the interpretation of the results.

\subsection{Local execution environment}

All measurements were collected in a controlled local environment, with the final dataset generated on bare metal rather than over a wide-area network. This design reduces network noise and allows certificate-path effects, chain exposure, and client/server cryptographic burden to be isolated more cleanly. It also leaves WAN latency, Internet path variability, queueing under production concurrency, and deployment-specific interactions with middleboxes and edge infrastructure outside the measurement setting.

Accordingly, the latency values reported here should be interpreted as comparative measurements within the tested stack, rather than as direct predictions of end-user Internet latency under arbitrary operating conditions.

\subsection{Single implementation stack}

The study is conducted on a specific implementation stack, namely OpenSSL~3 together with oqsprovider and liboqs. The results are implementation-dependent: different TLS libraries, certificate-handling paths, provider integrations, or post-quantum implementations could shift absolute values and potentially alter some lower-level trade-offs.

For that reason, the strongest claims of the paper are structural. Within a real-stack setting, hierarchy-sensitive signature placement can dominate the operational cost of post-quantum TLS authentication to an extent that flat algorithm comparisons fail to capture. The paper does not assert that every TLS implementation would produce exactly the same ratios or absolute timings.

\subsection{Unequal run counts and scope of inference}

Run counts were not uniform across scenarios. Fast scenarios were executed more extensively, whereas extremely heavy scenarios, especially those involving SLH-DSA in the leaf, were run fewer times because each handshake was much more expensive to execute.

A larger sample size improves the stability of descriptive estimates for a given scenario in the tested environment. It does not broaden the scope of the claims beyond that environment, nor does it remove implementation-specific behavior, measurement dependence across repeated runs, or other stack-level sources of bias.

The central findings of the paper do not depend on resolving small deltas among near-equivalent configurations. They rest on sharply separated operational regimes: sub-millisecond and low-millisecond all-ML or upper-layer-SLH cases on the one hand, and second-scale leaf-SLH cases on the other. Those separations remain visible in mean latency, p95 latency, transport volume, and server-side task-clock.

\subsection{Semantics of observed chain exposure}

A recurring methodological subtlety throughout the paper is that the effective chain observed by the client is not always equivalent to the naive logical PKI topology. In the dataset analyzed here, \texttt{chain\_len\_unique = 2} does not imply that the hierarchy is logically depth~2, and \texttt{served\_chain\_der\_bytes} does not carry fully homogeneous semantics across all topology classes.

This imposes interpretive discipline on chain-level analysis. For that reason, the paper relies primarily on \texttt{bytes\_read\_mean} and \texttt{chain\_bytes\_unique} for strict cross-depth transport reasoning, and treats \texttt{served\_chain\_der\_bytes} with explicit caution. As a result, some chain-level conclusions are stated in empirical terms tied to observed exposure, rather than in idealized PKI-topology terms.

\subsection{Perf as proxy rather than full function tracing}

The client/server performance-counter analysis is one of the strongest components of the paper, but it remains aggregate evidence rather than full internal function tracing. The results show clearly that leaf-SLH scenarios are overwhelmingly server-side and nearly compute-bound at the server, but they do not by themselves identify a unique internal function, call path, or provider-level routine as the exclusive source of cost. The performance-counter evidence should be interpreted as regime-level causal support rather than as proof of a uniquely identified internal mechanism.

\subsection{Economic model as derived interpretation}

The economic section of the paper is intentionally framed as an operational interpretation derived from measured server-side compute cost. The service classes, pricing assumptions, and cost-per-million-handshakes figures are illustrative and parameterized rather than tied to a specific commercial contract, cloud provider invoice, or production deployment profile. The resulting quantities therefore express the scale of the measured compute burden under the stated model, not the exact future bill of any particular organization.

\section{Conclusion}

\subsection{Main finding}

The main finding of this paper is that post-quantum migration in TLS~1.3 is best understood as a certificate-hierarchy design problem rather than as a flat comparison between signature algorithms. Across all four campaigns, the most important practical distinction is whether SLH-DSA reaches the handshake-exposed server leaf.

Once SLH-DSA is placed in that position, the authentication path enters a qualitatively different regime. Mean latency moves from the sub-millisecond or low-millisecond region into the $\sim$1.4-second range, server-side active compute becomes almost identical to end-to-end elapsed time, retained capacity collapses, and infrastructure multipliers rise by orders of magnitude. By contrast, scenarios that confine SLH-DSA to upper trust layers while preserving ML-DSA in the interactive leaf stay within a bounded and materially more plausible regime.

The paper therefore supports a precise conclusion. In TLS, the operational meaning of a post-quantum signature family is determined by its presence in the hierarchy and by its placement within the live authentication path.

\subsection{Implications for certificate hierarchy design}

This conclusion has direct implications for the design of post-quantum certificate hierarchies. A migration strategy should be evaluated in terms of standardization status, primitive-level conservatism, conceptual uniformity across the PKI, handshake exposure, and the location where cryptographic burden is paid.

The results suggest that uniform post-quantum placement across the entire hierarchy is not necessarily the most rational deployment strategy. The data support a more selective approach in which heavier signature families may be tolerated in upper trust layers while the interactive server leaf is protected from the regime-defining cost of SLH-DSA. Within the design space studied here, \texttt{root-SLH / intermediate-ML / leaf-ML} emerges as a materially penalized but still plausible strategy, whereas leaf-SLH configurations occupy a different operational class.

\subsection{What remains open}

Several lines of work remain open. First, the present study uses a single implementation stack. Repeating the same hierarchy-sensitive analysis across other TLS libraries and post-quantum integrations would strengthen confidence in the generality of the structural findings. Second, the present work focuses on full handshakes rather than resumed sessions. Session resumption may alter the practical weight of certificate-path cost in some deployment settings.

Finally, the broader post-quantum transition raises a more general design question that goes beyond the specific scenarios measured here: how should protocol designers and operators distribute cryptographic conservatism across the certification hierarchy when different positions in that hierarchy carry radically different operational meaning? The present paper does not settle that question in general, but it shows that answering it requires more than flat primitive comparison.

\printbibliography

@misc{rfc8446,
	author       = {Eric Rescorla},
	title        = {{The Transport Layer Security (TLS) Protocol Version 1.3}},
	series       = {Request for Comments},
	number       = {8446},
	howpublished = {RFC 8446},
	publisher    = {RFC Editor},
	year         = {2018},
	month        = aug,
	pagetotal    = {160},
	doi          = {10.17487/RFC8446},
	url          = {https://www.rfc-editor.org/info/rfc8446}
}

@techreport{fips203,
	author      = {{National Institute of Standards and Technology}},
	title       = {{Module-Lattice-Based Key-Encapsulation Mechanism Standard}},
	institution = {National Institute of Standards and Technology, U.S. Department of Commerce},
	type        = {Federal Information Processing Standards Publication},
	number      = {NIST FIPS 203},
	address     = {Gaithersburg, MD},
	year        = {2024},
	month       = aug,
	day         = {13},
	doi         = {10.6028/NIST.FIPS.203},
	url         = {https://doi.org/10.6028/NIST.FIPS.203}
}

@techreport{fips204,
	author      = {{National Institute of Standards and Technology}},
	title       = {{Module-Lattice-Based Digital Signature Standard}},
	institution = {National Institute of Standards and Technology, U.S. Department of Commerce},
	type        = {Federal Information Processing Standards Publication},
	number      = {NIST FIPS 204},
	address     = {Gaithersburg, MD},
	year        = {2024},
	month       = aug,
	day         = {13},
	doi         = {10.6028/NIST.FIPS.204},
	url         = {https://doi.org/10.6028/NIST.FIPS.204}
}

@techreport{fips205,
	author      = {{National Institute of Standards and Technology}},
	title       = {{Stateless Hash-Based Digital Signature Standard}},
	institution = {National Institute of Standards and Technology, U.S. Department of Commerce},
	type        = {Federal Information Processing Standards Publication},
	number      = {NIST FIPS 205},
	address     = {Gaithersburg, MD},
	year        = {2024},
	month       = aug,
	day         = {13},
	doi         = {10.6028/NIST.FIPS.205},
	url         = {https://doi.org/10.6028/NIST.FIPS.205}
}

@misc{oqs,
	author       = {{Open Quantum Safe Project}},
	title        = {{liboqs}},
	howpublished = {\url{https://openquantumsafe.org/liboqs/}},
	note         = {Accessed: 2026-05-19}
}

@misc{oqsprovider,
	author       = {{Open Quantum Safe Project}},
	title        = {{oqs-provider: Open Quantum Safe Provider for OpenSSL 3}},
	howpublished = {\url{https://github.com/open-quantum-safe/oqs-provider}},
	note         = {Accessed: 2026-05-19}
}

@inproceedings{sikeridis2020,
	author    = {Dimitrios Sikeridis and Panos Kampanakis and Michael Devetsikiotis},
	title     = {{Post-Quantum Authentication in TLS 1.3: A Performance Study}},
	booktitle = {Proceedings of the Network and Distributed System Security Symposium (NDSS 2020)},
	year      = {2020},
	month     = feb,
	doi       = {10.14722/ndss.2020.24203},
	url       = {https://www.ndss-symposium.org/wp-content/uploads/2020/02/24203.pdf}
}

@inproceedings{paul2021mixed,
	author    = {Sebastian Paul and Yulia Kuzovkova and Norman Lahr and Ruben Niederhagen},
	title     = {{Mixed Certificate Chains for the Transition to Post-Quantum Authentication in TLS 1.3}},
	booktitle = {{ASIA CCS '22: Proceedings of the 2022 ACM Asia Conference on Computer and Communications Security}},
	pages     = {727--740},
	publisher = {Association for Computing Machinery},
	year      = {2022},
	month     = may,
	day       = {30},
	doi       = {10.1145/3488932.3497755},
	url       = {https://doi.org/10.1145/3488932.3497755}
}

@article{montenegro2025framework,
	author    = {Jose A. Montenegro and Ruben Rios and Javier Lopez-Cerezo},
	title     = {{A Performance Evaluation Framework for Post-Quantum TLS}},
	journal   = {Future Generation Computer Systems},
	volume    = {175},
	number    = {108062},
	pages     = {1--14},
	publisher = {Elsevier},
	year      = {2026},
	month     = feb,
	doi       = {10.1016/j.future.2025.108062},
	issn      = {0167-739X},
	url       = {https://doi.org/10.1016/j.future.2025.108062}
}

@misc{rfc5280,
	author       = {Sharon Boeyen and Stefan Santesson and Tim Polk and Russ Housley and Stephen Farrell and David Cooper},
	title        = {{Internet X.509 Public Key Infrastructure Certificate and Certificate Revocation List (CRL) Profile}},
	series       = {Request for Comments},
	number       = {5280},
	howpublished = {RFC 5280},
	publisher    = {RFC Editor},
	year         = {2008},
	month        = may,
	pagetotal    = {151},
	doi          = {10.17487/RFC5280},
	url          = {https://www.rfc-editor.org/info/rfc5280}
}

@techreport{mlkemtlsdraft,
	author      = {Deirdre Connolly},
	title       = {{ML-KEM Post-Quantum Key Agreement for TLS 1.3}},
	institution = {Internet Engineering Task Force},
	publisher   = {Internet Engineering Task Force},
	type        = {Internet-Draft},
	number      = {draft-ietf-tls-mlkem-07},
	year        = {2026},
	month       = feb,
	day         = {12},
	pagetotal   = {11},
	note        = {Work in Progress},
	url         = {https://datatracker.ietf.org/doc/draft-ietf-tls-mlkem/07/}
}

@techreport{turner2026hybrid,
	author      = {Douglas Stebila and Scott Fluhrer and Shay Gueron},
	title       = {{Hybrid Key Exchange in TLS 1.3}},
	institution = {Internet Engineering Task Force},
	publisher   = {Internet Engineering Task Force},
	type        = {Internet-Draft},
	number      = {draft-ietf-tls-hybrid-design-16},
	year        = {2025},
	month       = sep,
	day         = {7},
	pagetotal   = {23},
	note        = {Work in Progress},
	url         = {https://datatracker.ietf.org/doc/draft-ietf-tls-hybrid-design/16/}
}

@techreport{uta2026pqcapp,
	author      = {Tirumaleswar Reddy.K and Hannes Tschofenig},
	title       = {{Post-Quantum Cryptography Recommendations for TLS-based Applications}},
	institution = {Internet Engineering Task Force},
	publisher   = {Internet Engineering Task Force},
	type        = {Internet-Draft},
	number      = {draft-ietf-uta-pqc-app-01},
	year        = {2026},
	month       = feb,
	day         = {24},
	pagetotal   = {23},
	note        = {Work in Progress},
	url         = {https://datatracker.ietf.org/doc/draft-ietf-uta-pqc-app/01/}
}

\newpage
\appendix

\section{Scenario Inventory}

This appendix provides the full experimental inventory used in the study. The main text discusses only those scenario distinctions needed for causal and operational interpretation, whereas this appendix preserves the broader inventory for traceability and reproducibility.

Scenario identifiers are constructed compositionally. They encode the key-establishment mode together with the signature-family placement across the logical certificate hierarchy. In the naming convention used throughout the paper, \texttt{root}, \texttt{int}, and \texttt{leaf} denote certificate positions in the hierarchy, while labels such as \texttt{ml} and \texttt{slh} denote the corresponding signature-family assignment. Depth-2 scenarios omit the intermediate position by construction.

\clearpage
\begin{landscape}
	\thispagestyle{plain}
	\vspace*{\fill}
	
	\begin{table}[H]
		\centering
		\caption{Experimental inventory and scenario matrix.}
		\label{tab:appendix_campaign_inventory_and_scenario_matrix}
		\begin{tabular}{llllllrr}
\toprule
Campaign & Scenario & TLS group & Root & Intermediate & Leaf & Depth & Runs  \\
\midrule
A & x25519\_\_leaf\_mldsa65 & x25519 & ML-DSA & — & ML-DSA & 2 & 10000  \\
A & x25519\_\_leaf\_slhdsashake192s & x25519 & SLH-DSA & — & SLH-DSA & 2 & 300  \\
A & x25519mlkem768\_\_leaf\_mldsa65 & x25519mlkem768 & ML-DSA & — & ML-DSA & 2 & 10000  \\
A & x25519mlkem768\_\_leaf\_slhdsashake192s & x25519mlkem768 & SLH-DSA & — & SLH-DSA & 2 & 300  \\
B & x25519mlkem768\_\_ml\_root\_\_ml\_int\_\_ml\_leaf & x25519mlkem768 & ML-DSA & ML-DSA & ML-DSA & 3 & 10000 \\
B & x25519mlkem768\_\_ml\_root\_\_ml\_int\_\_slh\_leaf & x25519mlkem768 & ML-DSA & ML-DSA & SLH-DSA & 3 & 300 \\
B & x25519mlkem768\_\_ml\_root\_\_slh\_int\_\_slh\_leaf & x25519mlkem768 & ML-DSA & SLH-DSA & SLH-DSA & 3 & 300  \\
B & x25519mlkem768\_\_slh\_root\_\_ml\_int\_\_ml\_leaf & x25519mlkem768 & SLH-DSA & ML-DSA & ML-DSA & 3 & 10000  \\
B & x25519mlkem768\_\_slh\_root\_\_ml\_int\_\_slh\_leaf & x25519mlkem768 & SLH-DSA & ML-DSA & SLH-DSA & 3 & 300  \\
B & x25519mlkem768\_\_slh\_root\_\_slh\_int\_\_slh\_leaf & x25519mlkem768 & SLH-DSA & SLH-DSA & SLH-DSA & 3 & 300  \\
C & x25519mlkem768\_\_ml\_root\_\_ml\_leaf & x25519mlkem768 & ML-DSA & — & ML-DSA & 2 & 10000  \\
C & x25519mlkem768\_\_ml\_root\_\_slh\_leaf & x25519mlkem768 & ML-DSA & — & SLH-DSA & 2 & 300  \\
C & x25519mlkem768\_\_slh\_root\_\_ml\_leaf & x25519mlkem768 & SLH-DSA & — & ML-DSA & 2 & 10000  \\
C & x25519mlkem768\_\_slh\_root\_\_slh\_leaf & x25519mlkem768 & SLH-DSA & — & SLH-DSA & 2 & 300  \\
D & mlkem768\_\_ml\_root\_\_ml\_leaf & mlkem768 & ML-DSA & — & ML-DSA & 2 & 10000 \\
D & mlkem768\_\_slh\_root\_\_ml\_int\_\_ml\_leaf & mlkem768 & SLH-DSA & ML-DSA & ML-DSA & 3 & 10000  \\
D & mlkem768\_\_slh\_root\_\_slh\_leaf & mlkem768 & SLH-DSA & — & SLH-DSA & 2 & 300  \\
\bottomrule
\end{tabular}

	\end{table}
	
	\vspace*{\fill}
\end{landscape}
\clearpage

\end{document}